\documentclass[acmlarge]{acmart}
\AtBeginDocument{%
  \providecommand\BibTeX{{%
    \normalfont B\kern-0.5em{\scshape i\kern-0.25em b}\kern-0.8em\TeX}}}

\setcopyright{acmcopyright}
\copyrightyear{2023}
\acmYear{2023}
\acmDOI{XXXXXXX.XXXXXXX}

\acmJournal{TOSEM}

\usepackage{listings}
\lstdefinelanguage
   [x64]{Assembler}     
   [x86masm]{Assembler} 
   {morekeywords={CDQE,CQO,CMPSQ,CMPXCHG16B,JRCXZ,LODSQ,MOVSXD, %
                  POPFQ,PUSHFQ,SCASQ,STOSQ,IRETQ,RDTSCP,SWAPGS, %
                  rax,rdx,rcx,rbx,rsi,rdi,rsp,rbp, %
                  r8,r8d,r8w,r8b,r9,r9d,r9w,r9b, %
                  r10,r10d,r10w,r10b,r11,r11d,r11w,r11b, %
                  r12,r12d,r12w,r12b,r13,r13d,r13w,r13b, %
                  r14,r14d,r14w,r14b,r15,r15d,r15w,r15b}} 
\usepackage{tikz}
\usepackage{tcolorbox}
\usepackage{amsmath}
\usepackage{xcolor}
\usepackage{multirow}
\usepackage{wrapfig}
\usepackage{subcaption}
\usepackage{colortbl}
\usepackage{tabulary}
\usepackage{tabularx}
\usepackage[linesnumbered,ruled]{algorithm2e}
\usepackage{cleveref}
\usepackage{pgfplots}
\usepackage[marginal]{footmisc}

\pgfplotsset{compat=1.16}
\crefname{section}{§}{§§}
\newcommand{\toolname}{\textsc{Asteria-Pro}}
\newcommand{\diaphora}{{\textit{Diaphora}}}
\newcommand{\etal}{\emph{et al.}}
\newcommand{\eg}{\emph{e.g.,}}

\newcommand{\gemini}{{\textit{Gemini}}}
\newcommand{\methodname}[1]{\textit{#1}}

\newcommand{\ysgbf}[1]{\textbf{#1}}
\newcommand{\siamese}[0]{Siamese architecture}
 \usepgfplotslibrary{statistics}
    \pgfplotsset{
        my boxplot style/.style={
            boxplot,
            draw=black,
            solid,
            fill=white,
            mark=*,
            every mark/.append style={
                fill=white,
            },
        },
    }

\begin{document}
\title{Asteria-Pro: Enhancing Deep-Learning Based Binary Code Similarity Detection by Incorporating Domain Knowledge}

\author{Shouguo Yang} 
\authornote{First Author and Second Author contribute equally to this work.}
\affiliation{%
  \institution{Institute of Information Engineering, Chinese Academy of Sciences}
  \city{Beijing}
  \country{China}
}
\affiliation{%
  \institution{School of Cyber Security, University of Chinese Academy of Sciences}
  \city{Beijing}
  \country{China}
}
\email{yangshouguo@iie.ac.cn}

\author{Chaopeng Dong}
\affiliation{%
  \institution{Institute of Information Engineering, Chinese Academy of Sciences}
  \city{Beijing}
  \country{China}
}
\affiliation{%
  \institution{School of Cyber Security, University of Chinese Academy of Sciences}
  \city{Beijing}
  \country{China}
}
\email{dongchaopeng@iie.ac.cn}
\authornotemark[1]

\author{Yang Xiao}
\affiliation{%
  \institution{Institute of Information Engineering, Chinese Academy of Sciences}
  \city{Beijing}
  \country{China}
}
\affiliation{%
  \institution{School of Cyber Security, University of Chinese Academy of Sciences}
  \city{Beijing}
  \country{China}
}
\email{xiaoyang@iie.ac.cn }
\authornote{Corresponding authors.}
\author{Yiran Cheng}
\affiliation{%
  \institution{Institute of Information Engineering, Chinese Academy of Sciences}
  \city{Beijing}
  \country{China}
}
\affiliation{%
  \institution{School of Cyber Security, University of Chinese Academy of Sciences}
  \city{Beijing}
  \country{China}
}
\email{chengyiran@iie.ac.cn}
\author{Zhiqiang Shi}
\email{shizhiqiang@iie.ac.cn}
\affiliation{%
  \institution{Institute of Information Engineering, Chinese Academy of Sciences}
  \city{Beijing}
  \country{China}
}
\affiliation{%
  \institution{School of Cyber Security, University of Chinese Academy of Sciences}
  \city{Beijing}
  \country{China}
}
\authornotemark[2]
\author{Zhi Li}
\affiliation{%
  \institution{Institute of Information Engineering, Chinese Academy of Sciences}
  \city{Beijing}
  \country{China}
}
\affiliation{%
  \institution{School of Cyber Security, University of Chinese Academy of Sciences}
  \city{Beijing}
  \country{China}
}
\email{lizhi@iie.ac.cn}
\author{Limin Sun}
\affiliation{%
  \institution{Institute of Information Engineering, Chinese Academy of Sciences}
  \city{Beijing}
  \country{China}
}
\affiliation{%
  \institution{School of Cyber Security, University of Chinese Academy of Sciences}
  \city{Beijing}
  \country{China}
}
\email{sunlimin@iie.ac.cn}
\begin{abstract}
The widespread code reuse allows vulnerabilities to proliferate among a vast variety of firmware.
There is an urgent need to detect these vulnerable code effectively and efficiently.
By measuring code similarities, \textit{AI-based binary code similarity detection} is applied to detecting vulnerable code at scale.
Existing studies have proposed various function features to capture the commonality for similarity detection.
Nevertheless, the significant code syntactic variability induced by the diversity of IoT hardware architectures diminishes the accuracy of binary code similarity detection.
In our earlier study and the tool \methodname{Asteria}, we adopted a Tree-LSTM network to summarize function semantics as function commonality, and the evaluation result indicates an advanced performance.
However, it still has utility concerns due to excessive time costs and inadequate precision while searching for large-scale firmware bugs.

To this end, we propose a novel deep learning enhancement architecture by incorporating domain knowledge-based pre-filtration and re-ranking modules, and develop a prototype named \toolname{} based on \methodname{Asteria}.
The pre-filtration module eliminates dissimilar functions, thus reducing the subsequent deep learning model calculations. 
The re-ranking module boosts the rankings of vulnerable functions among candidates generated by the deep learning model.
Our evaluation indicates that the pre-filtration module cuts the calculation time by 96.9\%, and the re-ranking module improves MRR and Recall by 23.71\% and 36.4\%, respectively.
By incorporating these modules, \toolname{} outperforms existing state-of-the-art approaches in the bug search task by a significant margin.
Furthermore, our evaluation shows that embedding baseline methods with pre-filtration and re-ranking modules significantly improves their precision.
We conduct a large-scale real-world firmware bug search, and \toolname{} manages to detect 1,482 vulnerable functions with a high precision 91.65\%.
\end{abstract}



\maketitle

\section{Introduction}
Code reuse is very popular in IoT firmware to facilitate its development~\cite{woo2021v0finder}. 
Unfortunately, code reuse also introduces vulnerabilities concealed in the original code into a variety of firmware~\cite{cui2013firmware}.
The security and privacy of our lives are seriously threatened by the widespread use of these firmware~\cite{Wurm:DAC:2016}.
Even though the vulnerabilities have been publicly disclosed, there are a large number of firmware versions that still contain them due to delayed code upgrades or code compatibility issues~\cite{bogart2021and}.
Recurring vulnerabilities, often referred to as ``N-day vulnerabilities'', cannot be detected through symbol information such as function names because this type of information is usually removed during firmware compilation. Additionally, the source code of firmware is typically unavailable as IoT vendors only provide binary versions of their firmware.

To this end, binary code similarity detection (BCSD) is applied to quickly find homologous vulnerabilities in a large amount of firmware~\cite{david2018firmup}.
The BCSD technique focuses on determining the similarity between two binary code pieces.
As for the vulnerability search, BCSD looks for other vulnerable functions that are similar to one that is already known to be vulnerable.
In addition to the vulnerability search, BCSD has been widely used for other security applications such as code plagiarism detection~\cite{Basit:2005, schulman2005finding, luo2014semantics}, malware detection~\cite{hu2009large,hu2013mutantx}, and patch analysis~\cite{gao2008binhunt, wang2000bmat, dullien2005graph}. 
Despite many existing research efforts, the diversity of IoT hardware architectures and software platforms poses challenges to BCSD for IoT firmware. 
There are many different instruction set architectures (ISA) for IoT firmware, such as ARM, PowerPC, X64, and X86.
The instructions are different, and the rules, such as the calling convention and the stack layout, also differ across different ISAs.
It is non-trivial to find homologous vulnerable functions across various architectures.

BCSD methods can be generally classified into two categories: i) dynamic analysis-based methods and ii) static analysis-based methods.
The methods based on dynamic analysis capture the runtime behavior as function semantic features by running target functions, where the function features can be I/O pairs of function~\cite{pewny2015cross} or system calls during program execution \cite{egele2014blanket}, etc.
They are not scalable for large-scale firmware analysis since running firmware requires specific devices and emulating firmware is also difficult~\cite{zaddach2014avatar, gustafson2019toward, chen2016towards}.
The methods based on static analysis mainly extract statistical features from assembly code.
An intuitive way is to calculate the edit distance between assembly code sequences~\cite{David:Tracelet}.
They cannot be directly applied across architectures since instruction sets are totally distinct.
Architecture-independent statistical features of functions are proposed for similarity detection~\cite{eschweiler2016discovre}.
These features are less affected across architectures such as the number of function calls, strings, and constants.
Furthermore, the control flow graph (CFG) at the assembly code level is utilized by conducting a graph isomorphism comparison for improving the similarity detection ~\cite{eschweiler2016discovre, feng2016scalable}.
Based on statistical features and CFG, Gemini~\cite{xu2017neural} leverages the graph embedding network to encode functions as vectors for similarity detection.
With the application of deep learning models in programming language analysis, various methods have recently appeared to employ such models to encode binary functions in different forms and calculate function similarity based on function encoding~\cite{liu2018alphadiff, 280046, trex, wang2022jtrans}.
Static analysis-based methods are faster and more scalable for large-scale firmware analysis but often produce false positives due to the lack of semantic information.
Since homologous vulnerable functions in different architectures usually have the same semantics, a cross-architecture BCSD should be able to capture the semantic information about functions in a way that can be scaled. 

In our previous work \methodname{Asteria}~\cite{yang2021asteria}, we first utilized the Tree-LSTM network to encode the AST in an effort to capture its semantic representation.
In particular, Tree-LSTM is trained using a \methodname{siamese}~\cite{he2018twofold} architecture to understand the semantic representation by feeding homologous and non-homologous function pairs into the Tree-LSTM network. 
Consequently, the Tree-LSTM network learns function semantic representations to distinguish between homologous and non-homologous functions.
To further improve the accuracy, we also use the call graph to calibrate the AST similarity.
Precisely, we count callee functions of target functions in the call graph to measure the difference in function calls.
The final function similarity is determined by calibrating the AST similarity with the disparity in function calls.
In our previous evaluation, \methodname{Asteria} outperformed the available state-of-the-art methods, \methodname{Gemini} and \methodname{Diaphora}, in terms of accuracy.
The evaluation results demonstrate the superiority of function semantic extraction by encoding AST with the Tree-LSTM model.
However, encoding the AST incurs a clear temporal cost for \methodname{Asteria}.
According to our earlier research~\cite{yang2021asteria}, the entire AST encoding process takes about one second.
When \methodname{Asteria} is applied to vulnerability detection, where there are numerous functions to perform similarity calculations given a vulnerable function, the time cost becomes unacceptable.
Since the majority of candidate functions are non-homologous, there is room for enhancing the efficiency of \methodname{Asteria}.
In other words, non-homologous candidate functions differ from vulnerable functions in certain characteristics that we can exploit to skip the majority of non-homologous functions more effectively.
In addition, the evaluations do not align with the approaches used in the majority of real-world vulnerability detection efforts~\cite{xu2017neural, feng2016scalable, massarelli2019safe, zuo2018neural, li2021palmtree}, including our prior study \methodname{Asteria}.
Vulnerability detection involves retrieving homologous (vulnerable) functions from a large pool of functions.
Consequently, their performance in detecting vulnerabilities is insufficiently described.
It is necessary to evaluate the performance of \methodname{Asteria} on the vulnerability search task.
Moreover, according to the result in the real world vulnerability detection~\cite{yang2021asteria}, \methodname{Asteria} suffers from high false positives, which affects its effectiveness in reality.

There are two main challenges that hinder \methodname{Asteria} from being practical for large-scale vulnerability detection:
\begin{itemize}
    \item \textbf{Challenge 1 (C1).} It's challenging to filter out the majority of non-homologous functions before encoding ASTs, while retaining the homologous ones, to speed up the vulnerability-detection process.
    
    \item \textbf{Challenge 2 (C2).} It's challenging to distinguish similar but non-homologous functions. Despite \methodname{Asteria}'s high precision in homologous and non-homologous classification, it still yields false positives when distinguishing functions with similar ASTs.
\end{itemize}

We design \toolname{} by introducing domain knowledge as two answers, \textbf{A1} and \textbf{A2} to overcome these two challenges.
Our fundamental concept is that introducing inter-functional domain knowledge will helps \toolname{} achieve greater precision combined the intra-functional semantic knowledge deep learning model learned.
\toolname{} consists of three modules: \textbf{1)} domain knowledge-based (DK-based) pre-filtration, \textbf{2)} deep learning-based (DL-based) similarity detection, and \textbf{3)} DK-based re-ranking, among them DL-based similarity detection is basically based on \methodname{Asteria}.
Domain knowledge is fully exploited for different purposes in DK-based pre-filtration and re-ranking.
In pre-filtration module, \toolname{} aims to skip as many as possible non-homologous function by comparing lightweight robust features (\textbf{A1}). 
Meanwhile, filtration is required to retain all homologous functions.
To this end, we conducted a preliminary study into the filtering performance of several lightweight function features.
According to the findings of the study, we propose a novel algorithm that successfully employs three distinct function features in the filter.
In the re-ranking module, \toolname{} confirms the homology of functions by comparing call relationships (\textbf{A2}), based on the assumption that functions designed for distinct purposes have different call relationships.

Our evaluation indicates that \toolname{} significantly outperforms existing state-of-the-art methods in terms of both accuracy and efficiency.
Compared with \methodname{Asteria}, \toolname{} successfully cuts the detection time of \methodname{Asteria} by \textbf{96.90\%} by incorporating DK-based pre-filtration module.
In the vulnerability-search task, \toolname{} has the shortest average search time than other baseline methods.
By incorporating DK-based re-ranking, \toolname{} manages to enhance the MRR and Recall@Top-1 by 23.71\% and 36.4\%, to \textbf{90.8\%} and \textbf{89.6\%}, respectively.
We have also applied our enhancement framework to embed baseline methods, and the evaluation results demonstrate a significant improvement in the precision of these methods.
\toolname{} identifies 1,482 vulnerable functions with a high precision of 91.65\% by conducting a large-scale real-world firmware vulnerability detection utilizing 90 CVEs.
Moreover, the detection results of CVE-2017-13001 demonstrate that \toolname{} has an advanced capacity to detect inlined vulnerable code.


Our contributions are summarized as follows:
\begin{itemize}
    \item  We conduct a preliminary study to demonstrate the effectiveness of various simple function features in identifying non-homologous functions.
    \item  To the best of our knowledge, it is the first work to propose incorporating domain knowledge before and after deep learning models for vulnerability detection optimization. We implement the domain knowledge-based pre-filtration and re-ranking algorithms and equip Asteria with them.
    \item The evaluation indicates the pre-filtration module significantly reduces the detection time, and re-ranking module improves the detection precision by a fairly amount.
    The \toolname{} outperforms existing state-of-the-art methods in terms of both accuracy and efficiency. In evaluation~\ref{sec:rq1_eval}, we find that the performance of distinct BCSD methods may vary widely in different usage scenarios.
    \item We demonstrate the utility of \toolname{} by conducting a large-scale, real-world firmware vulnerability detection. \toolname{} manages to find 1,482 vulnerable functions with a high precision of 91.65\%. We analyze the vulnerability distribution in widely-used software from various IoT vendors to illustrate our inspiring findings. 
\end{itemize}
\section{Background}
We first briefly describe the AST structure adopted in this work, followed by a demonstration of the AST holding a more stable structure than CFG across architectures.
Then we introduce the Tree-LSTM model utilized in AST encoding.
Finally, the broad problem definition for the application of BCSD to bug search is given.



\subsection{Abstract Syntax Tree}

\begin{table}[!h]
    \caption{Statements and Expressions in ASTs. We count the statements and expressions for nodes in ASTs after the decompilation by \texttt{IDA Pro} and list the common statements and expressions. This table can be extended if new statements or expressions are introduced.}
    \label{tab:mapping}
    \centering
    \resizebox{15cm}{!}{
    \begin{tabularx}{\linewidth}{c|ccX}
        \hline
         &   Node Type & Label & Note  \\ 
         \hline
         \multirow{9}*{Statement} 
         ~& if   & 1 & if statement   \\
         ~& block   & 2& instructions executed sequentially\\
         ~& for   & 3 & for loop statement\\
         ~& while   & 4 & while loop statement\\
         ~& switch   & 5 & switch statement\\
         ~& return   & 6 &return statement \\
         ~& goto   & 7 & unconditional jump \\
         ~& continue   & 8 &continue statement in a loop\\
         ~& break   & 9 &break statement in a loop \\
         \hline
         \multirow{8}*{Expression}
            & \multirow{2}*{asgs} & \multirow{2}*{10$\sim$17} & assignments, including assignment, assignment after or, xor, and, add, sub, mul, div\\
             & \multirow{2}*{cmps} & \multirow{2}*{18$\sim$23}  & comparisons including equal, not equal, greater than, less than, greater than or equal to, and less than or equal to.\\
             & \multirow{2}*{ariths} & \multirow{2}*{24$\sim$34} & arithmetic operations including or, xor, addition, subtraction, multiplication, division, not, post-increase, post-decrease, pre-increase, and pre-decrease \\
             & \multirow{2}*{other} & \multirow{2}*{34$\sim$43} & others including indexing, variable, number, function call, string, asm, and so on.\\
         \hline 
    \end{tabularx}
    }
\end{table}

\begin{figure}[h]
    \centering
    \includegraphics[scale=0.4]{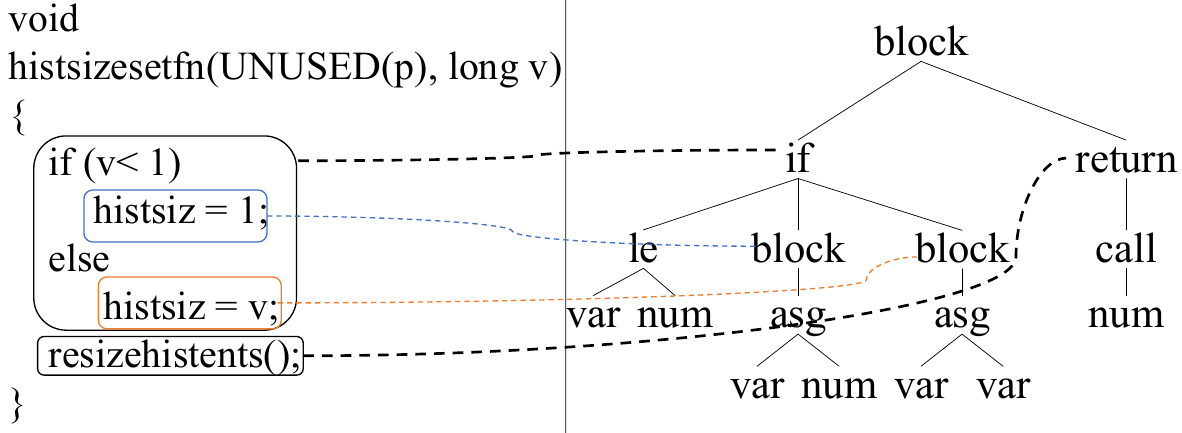}
    \caption{Source code of function \textit{histsizesetfn} and the corresponding decompiled AST of x86 architecture. }
    \label{fig:source_to_ast}
\end{figure}

\subsubsection{AST Description} \label{sec:ast_intro}
An AST is a tree representation of the abstract syntactic structure of code in the compilation and decompilation processes.
This work focuses on the ASTs extracted by decompiling binary functions.
Different subtrees in an AST correspond to different code scopes in the source code.
Figure \ref{fig:source_to_ast} shows a decompiled AST corresponding to the source code of function \texttt{histsizesetfn} in \texttt{zsh v5.6.2} on the left. 
The \texttt{zsh} is a popular shell software designed for interactive use, and the function \texttt{histsizesetfn} sets the value of a parameter.
The lines connecting the source code and AST in Figure~\ref{fig:source_to_ast} show that {a node} in the AST corresponds to an expression or a statement in the source code.
A variable or a constant value is represented by a leaf node in AST.
We group nodes in an AST into two categories: i) statement nodes and ii) expression nodes according to their functionalities shown in Table~\ref{tab:mapping}.
Statement nodes control the function execution flow while expression nodes perform various calculations.
Statement nodes include \textit{if, for, while, return, break} and so on.
Expression nodes include common arithmetic operations and bit operations.

\begin{figure}[tbh]
\centering
    \begin{minipage}[t]{0.48\linewidth}
    \centering
    \includegraphics[scale=0.35]{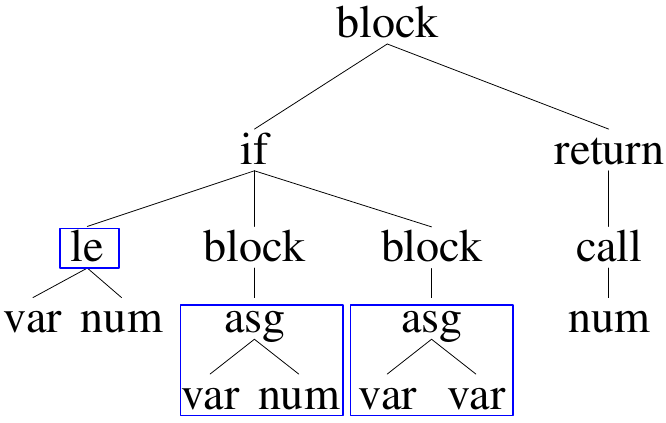}
    \end{minipage}
    \begin{minipage}[t]{0.48\linewidth}
    \centering
    \includegraphics[scale=0.35]{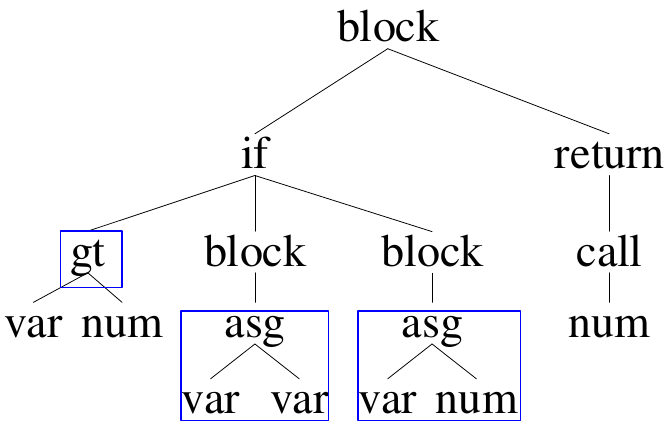}
    \end{minipage}
    \begin{minipage}[t]{0.48\linewidth}
    \subcaption{AST for x86 platform}
    \end{minipage}
    \begin{minipage}[t]{0.48\linewidth}
    \subcaption{AST for ARM platform}
    \end{minipage}
    \begin{minipage}[t]{0.48\linewidth}
    \centering
    \includegraphics[scale=0.29]{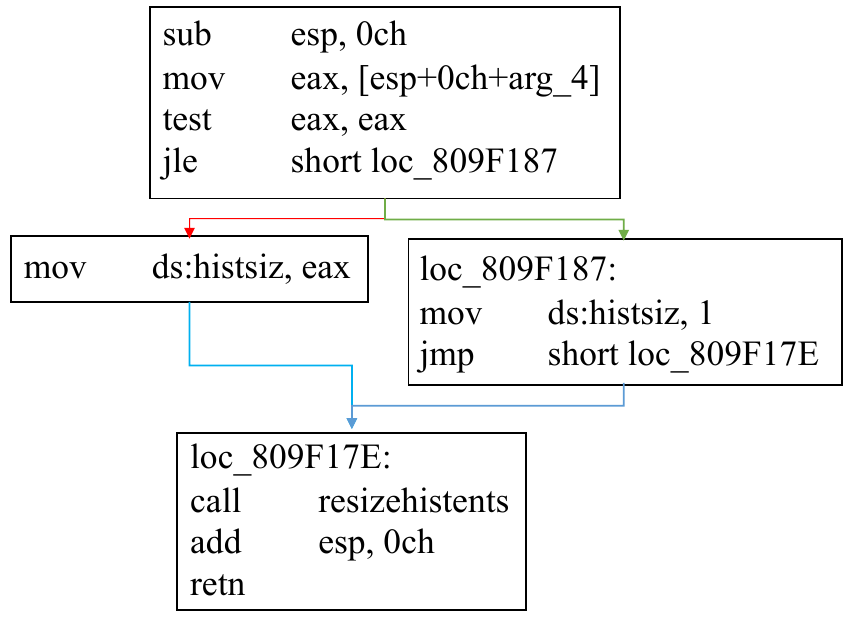}
    \end{minipage}
     \begin{minipage}[t]{0.48\linewidth}
    \centering
    \includegraphics[scale=0.35]{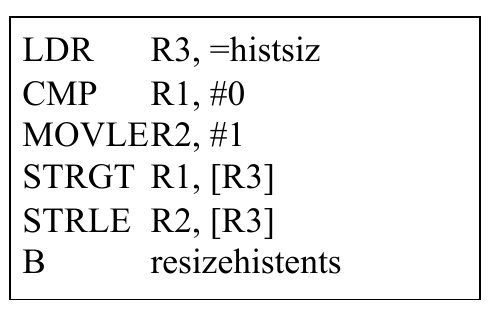}
    \end{minipage}
    \begin{minipage}[t]{0.48\linewidth}
    \subcaption{CFG for x86 platform}
    \end{minipage}
    \hfill
    \begin{minipage}[t]{0.48\textwidth}
    \subcaption{CFG for ARM platform}
    \end{minipage} 
    \caption{ASTs and CFGs of the function \textit{histsizesetfn} under different architectures.}
    \label{fig:comparisonofcfg}
\end{figure}

\subsubsection{AST Structure Superiority} \label{sec:astvscfg}
Both CFG and AST are structural representations of a function.
The CFG of a function contains the jump relationships between basic blocks that contain straight-line code sequences~\cite{hennessy2011computer}.
Though CFG has been used for similarity measurement in BCSD~\cite{eschweiler2016discovre}, David {\etal}~\cite{David:Tracelet} demonstrated that CFG structures differ significantly across different architectures. 
We observe that AST shows better architectural stability across architectures compared with CFG.
It is because AST is generated from the machine-independent intermediate presentations, which are disassembled from assemble instructions during the decompilation process~\cite{cifuentes1995decompilation}.
Figure~\ref{fig:comparisonofcfg} depicts the evolution of ASTs and CFGs for the x86 and ARM architectures, respectively.
For the CFGs from x86 to ARM, we observe that the number of basic blocks changes from 4 to 1, and the number of assembly instructions has changed a lot.
However, the ASTs, which are based on a higher-level intermediate representation, differ very slightly between x86 and ARM, with the differences highlighted by blue boxes.
In addition, AST maintains the semantics of functionality, making it an ideal structure for cross-platform similarity detection.

\subsection{Tree-LSTM Model}
In natural language processing, Recursive Neural Networks (RNN) are widely applied and perform better than Convolutional Neural Networks~\cite{yin2017comparative}.
RNNs take sequences of arbitrary lengths as inputs considering that a sentence can consist of any number of words.
However, standard RNNs are not capable of handling long-term dependencies due to the gradient vanishing and gradient exploding problems.
As one of the variants of RNN, Long Short-Term Memory (LSTM)~\cite{HochreiterLong} has been proposed to solve such problems. 
LSTM introduces a gate mechanism including the input, forget, and output gates. 
The gates control the information transfer to avoid the gradient vanishing and exploding (calculation details in Section~\cref{sec:treelstm}).
Nevertheless, LSTM can only process sequence input but not structured input.
Tree-LSTM is proposed to process tree-structured inputs~\cite{tai2015improved}.
The calculation by Tree-LSTM model is from the bottom up.
For each non-leaf node in the tree, all information from child nodes is gathered and used for the calculation of the current node.
In sentiment classification and semantic relatedness tasks, Tree-LSTM performs better than a plain LSTM structure network.
There are two types of Tree-LSTM proposed in the work \cite{treelstm}: Child-Sum Tree-LSTM and Binary Tree-LSTM.
Researchers have shown that Binary Tree-LSTM performs better than Child-Sum Tree-LSTM~\cite{treelstm}.
Since the Child-Sum Tree-LSTM does not take into account the order of child nodes, while the order of statements in AST reflects the function semantics, we use the Binary Tree-LSTM for our AST encoding.

\subsection{Function Call Compilation Optimization}\label{sec:func_call_opt}
There are two main types of function call compilation optimization that can impact binary code similarity analysis: function inline and intrinsic functions.

\noindent
\textbf{Function inline.} Function inline is a compiler optimization technique where the code of a called function is inserted directly into the calling function, rather than making a separate function call. This can improve program performance by reducing the overhead of function calls and improving cache utilization. The decision to inline a function is typically made by the compiler based on various factors such as function size, frequency of calls, and available register space.

\noindent
\textbf{Intrinsic function.} Intrinsic functions (also known as built-in functions) are special functions that are implemented by the compiler itself and are mapped to a single instruction or a sequence of instructions in the target architecture. These functions provide low-level access to the hardware and are used to implement various low-level operations, such as arithmetic, bit manipulation, and memory access. Intrinsic functions are often used in performance-critical code, where the use of low-level instructions can lead to significant speedups compared to equivalent code written in a higher-level language.

\section{Preliminary Study}~\label{sec:preliminary_study}
This study aims to assess and uncover accessible function features that are effective at identifying non-homologous functions to guide our pre-filtration design.
To evaluate the features, we prepare the code base and incorporate a number of metrics (\cref{sec:benchmark}).
We focus primarily on evaluating and comparing prevalent conventional features present in existing remarkable works (\cref{sec:feature_eval}).

\subsection{Evaluation Benchmark} \label{sec:benchmark}
\subsubsection{Dataset} \label{sec:codebase}
To derive robust features, we compile a large collection of binaries from 184 open source software (OSS), including widely used OpenSSL, FFmpeg, Binutils, etc.
Since our tool aims to conduct similarity detection across different architectures, we compile these open source software for four common architectures: X86, X64, ARM, and PowerPC.
In addition, we align the default compilation settings during compilation with real-world usage. 
After compilation, numerous test binaries with ``test'' or ``buildtest'' as a prefix or suffix are generated to test the software's functionality.
These test binaries are removed from the collection because 1) their functions are simple and comprise only a few lines of code. 2) do not participate in the real execution of software function.
After removal, the binary collection retains 1,130 binaries, or 226 for each architecture.

We create a large dataset consisting of pairs of homologous and non-homologous functions based on their function names. 
Function names are retained in the software after compilation, allowing us to construct the dataset.
To create homologous function pairs, we select binary functions with the same function names within the same software.
On the other hand, functions with different names were considered non-homologous.
For example, if function $F$ is present in the source code, compilation would generate four versions of binary functions for different instruction set architectures: $F_{x86}$, $F_{x64}$, $F_{arm}$, and $F_{ppc}$. 
These variants of functions are considered homologous to each other. 
We extract a total of 529,096 binary functions, comprising 132,274 unique functions for each architecture. 
To avoid overfitting in final evaluation, we randomly selected 40,111 functions from each architecture. 
Among them, we randomly chose $n$ functions as source functions to evaluate the filtering capability of diverse features.
For each source function $F^{B}{X}$, we constructed a pool of candidate functions consisting of $M$ randomly selected binary functions and three homologous functions of $F^{B}_{X}$. As a result, each source function $F^{B}_{X}$ forms three homologous pairs and $M$ non-homologous pairs.

\subsubsection{Metrics} \label{sec:metrics}
True positive rate (TPR) and false positive rate (FPR) are utilized to evaluate the filtering capability of various features.
TPR demonstrates the feature's capacity to retain homologous functions, while FPR demonstrates its capacity to exclude non-homologous functions.
In the subsequent filtering phase, our goal is \textit{to identify features that can filter out non-homologous functions as effectively as possible (low FPR) while maintaining all homologous functions (very high TPR).}

For a source function $F^{B}_{X}$, all function pairs in candidate function pool are measured by various feature similarity scores.
The function pairs with similarity scores below a threshold value $T$ are filtered.
In the remaining function pairs, the homologous function pairs are regarded as true positives $TP$ while the non-homologous function pairs are regarded as false positives $FP$.
The following equations illustrate how we calculate these three metrics for various features:
\begin{gather} \label{eq:recall_fp}
	TPR = \frac{\sum_{i=1}^nTP_i^p}{3 \times n} \\ 
    FPR =  \frac{\sum_{i=1}^nFP_i}{n \times (M \times 4-4)}
\end{gather}


\subsection{Candidate Features Evaluation} \label{sec:feature_eval} 
We aim to identify the most efficient and effective filter features by evaluating existing features proposed in previous studies and their variants. Based on the evaluation results, we select and improve candidate features to meet the filter requirements, which is to remove as many non-homologous functions as possible while retaining all homologous ones.

\subsubsection{Feature Selection}
We gather basic features from prior research~\cite{yang2021asteria,eschweiler2016discovre,xue2018accurate, xu2017neural} and categorize them into two groups: CFG-family features and AST-family features. 

The CFG-family features include four types of numeric features: the number of instructions (\textit{No. Instruction}), arithmetic instructions (\textit{No. Arithmetic}), call instructions (\textit{No. Callee}), and logical instructions (\textit{No. Logic}), along with two constant features: string constants (\textit{String Constant}) and numeric constants (\textit{Numeric Constant})~\cite{eschweiler2016discovre}. We also introduce a newly proposed feature called the \textit{named callee list} (\textit{NCL}) to capture the text sequence information of callee functions that retain their function names due to dynamic linking. In particular, \textit{NCL} is designed to be a list of callee functions that are either imported or exported functions. These functions retain their original names as they are used as identifiers to reference the functions in other parts of the code.

Since AST is necessary for model encoding calculation (\cref{sec:func_similarity}), we summarize three syntactic features as AST-family features:
\begin{itemize}
\item \textit{No. AST Nodes}: The number of AST nodes.
\item \textit{AST Node Cluster}: The number of different node types in the AST. For example, in Figure~\ref{fig:source_to_ast}, the AST node cluster is denoted as $[block: 3, if: 1, return: 1, call: 1, num: 3, block: 2, asg: 2, var: 4, le: 1]$.
\item \textit{AST Fuzzy Hash}: We first generate a node sequence by traversing the AST preorder. Then we apply the fuzzy hash algorithm~\cite{lee2017comparison} to generate the fuzzy hash of the AST.
\end{itemize}


\subsubsection{Feature Similarity Calculation}
 The format of features divides them into two types with distinct similarity calculations: value type and sequence type.
Value type features consist of \textit{No. Instruction}, \textit{No. Arithmetic}, \textit{No. Logic}, \textit{No. Callee}, and \textit{No. AST nodes}.
Sequence type features consist of \textit{Numeric Constant}, \textit{String Constant}, \textit{AST Node Cluster}, \textit{AST Fuzzy Hash}, and \textit{NCL}.
For value type features, we use the relative difference ratio ($RDR$) as shown below for similarity calculation:
\begin{equation}\label{eq:rdr}
    RDR(V_1, V_2) = 1-\frac{ abs( V_1 - V_2)}{max(V_1, V_2)}
\end{equation}
\noindent where $V_1, V_2$ are feature values.
For each sequence-type feature, we first sort the feature's items and then concatenate them into a single sequence.
Then, we employ the common sequence ratio (CSR) based on the longest common sequence (LCS) as follows:
\begin{equation}\label{eq:csr}
    CSR(S_1, S_2) = \frac{2 \times LCS( S_1, S_2)}{len(S_1) + len(S_2)}
\end{equation}
where $S_1, S_2$ are feature sequences, and function $LCS(\cdot, \cdot)$ returns the length of the longest common sequence between $S_1, S_2$.
The above two equations are used for similarity calculation of various features.
\begin{figure}[t]
    \centering
    \begin{minipage}{0.31\textwidth}
        \centering
    \includegraphics[width=\linewidth]{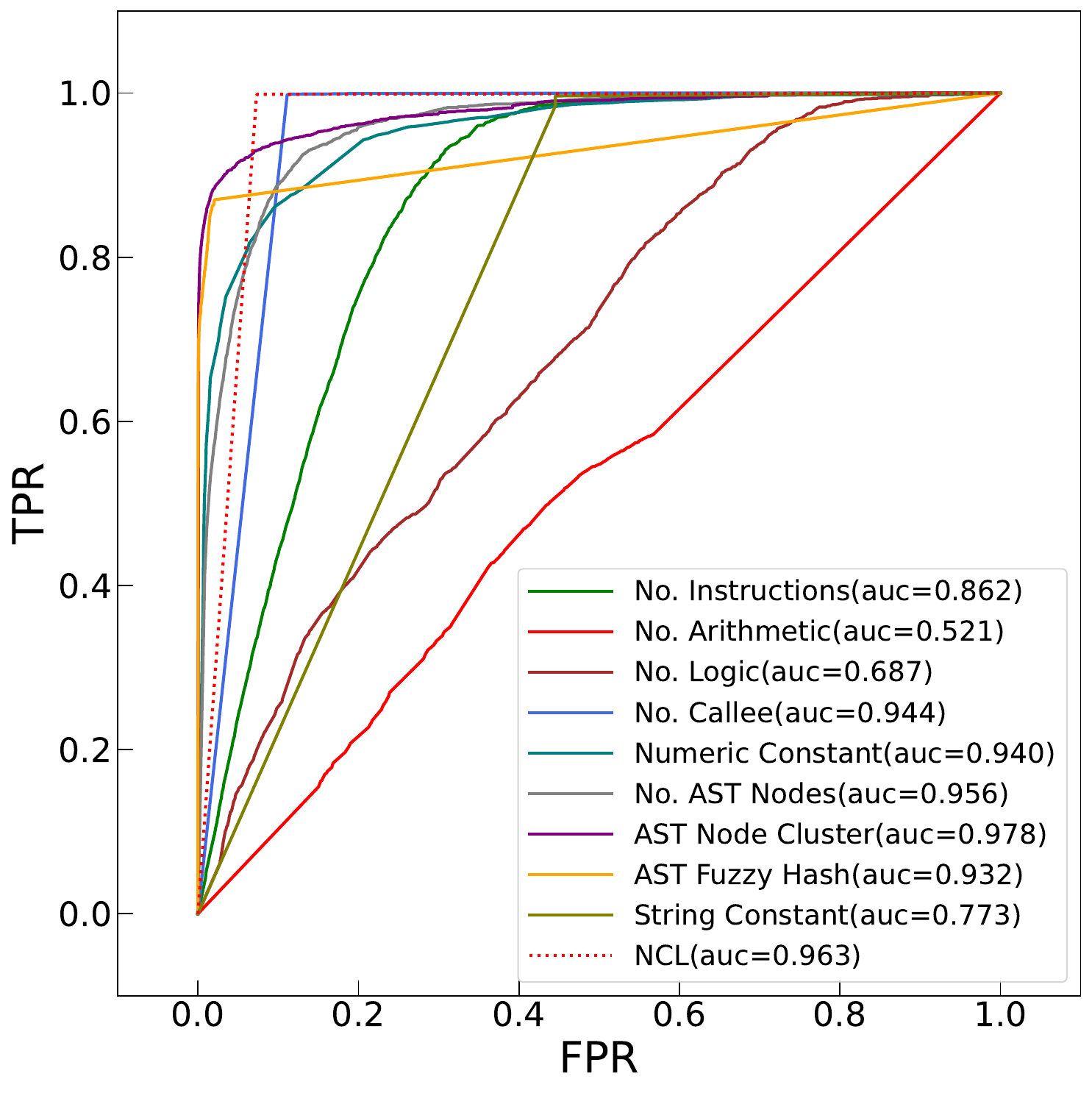}
    \end{minipage}
        \begin{minipage}{0.31\textwidth}
        \centering
    \includegraphics[width=\linewidth]{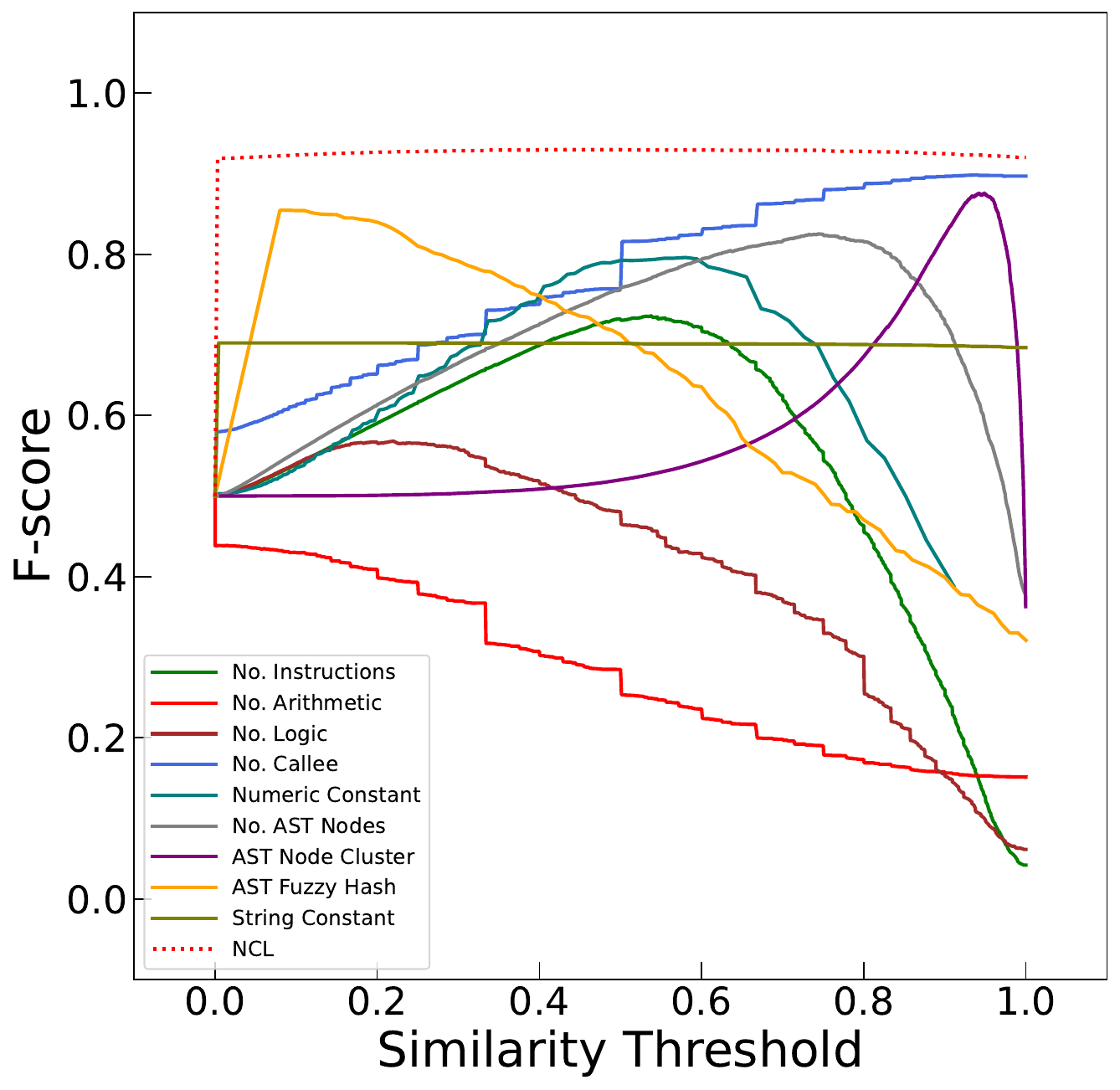}
    \end{minipage}
    \begin{minipage}{0.36\textwidth}
        \centering
    \includegraphics[width=\linewidth]{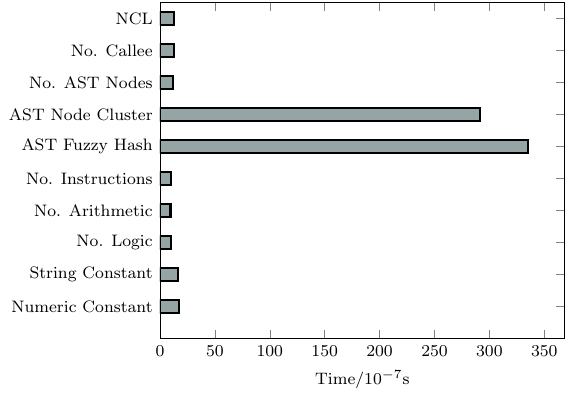}
    \end{minipage}

    \begin{minipage}{0.33\textwidth}
        \caption{ROC Curves for Features.}
    \label{fig:roc_old_fea}
    \end{minipage}
       \begin{minipage}{0.33\textwidth}
       \caption{$F_{score}$ of Features.}
    \label{fig:fscore_old_fea}
    \end{minipage}
     \begin{minipage}{0.33\textwidth}
        \caption{Time Costs of Similarity Calculation for Different Features.}
    \label{fig:fea_time_cost}
    \end{minipage}
  
\end{figure}

\subsubsection{Evaluation Results.}
In the evaluation, the values for $n$ and $M$ in \cref{sec:metrics} are set to $1000$ and $20,000$, respectively.
As depicted in Figure~\ref{fig:roc_old_fea}, TPRs and FPRs calculated for each feature under various thresholds are presented as a receiver operating characteristic (ROC)~\cite{zweig1993receiver} curve.
Additionally, we compute the area under the ROC curve (AUC), which reflects the feature's ability to distinguish between homologous and non-homologous functions.
The AUC values of the features extracted from AST (i.e., \textit{No. AST Nodes}, \textit{AST Node Cluster}, and \textit{AST Fuzzy Hash}) are high, as presented in the Figure~\ref{fig:roc_old_fea}.
However, when the TPR is high, they generate a high FPR.
Figure~\ref{fig:fea_time_cost} depicts the time costs associated with similarity calculations for various features. 
Clearly, sequence type features require more time than value type features.
Nonetheless, their time consumption falls within an acceptable range of magnitudes. At least $10^5$ exact calculations can be completed every second.

We observe in Figure~\ref{fig:roc_old_fea} that at high TPR (0.996), the \textit{No. Callee} feature produces a relatively lower FPR (0.111). Recalling the requirement of the filtering phase, we aim to select features with a low FPR at a very high TPR. Features with high AUC do not necessarily meet our objective. For example, the feature \textit{AST Node Cluster} has a higher FPR (0.47) than the feature \textit{No. Callee} (FPR = 0.111) under the same TPR (0.996), even though the feature \textit{AST Node Cluster} has a higher AUC (0.978) than the feature \textit{No. Callee} (AUC = 0.944).
In this regard, we propose a new metric, $F_{score}$, which indicates a high TPR and a lower FPR.
\begin{equation}
    F_{score} = \frac{1}{\frac{1}{TPR} + FPR}
    \label{eq:fscore}
\end{equation}
Figure~\ref{fig:fscore_old_fea} plots the $F_{score}$ curves of various features at different similarity thresholds. The results indicate that the "\textit{NCL}" feature has the highest $F_{score}$ of 0.92 among all the candidate features. It achieves a high true positive rate at a low false positive rate, with a relatively high AUC score of 0.963.
The "\textit{No. Callee}" feature performs slightly worse, with an AUC score of 0.944 and an $F_{score}$ of 0.902.
The "\textit{String Constant}" feature shows a relatively high $F_{score}$ at a very low threshold (e.g., 0.01) since it decisively determines the homology of functions. In particular, if two functions have the same strings, they are highly likely to be homologous. Although the $F_{score}$ does not increase as the threshold increases, it is because some functions do not include string constants, which limits the number of true positive pairs.
Based on the filtering performance of the candidate features, we have decided to use \textit{NCL} along with \textit{No. Callee} and \textit{String Constant} for our prefiltering design.
\begin{figure}[t]
    \centering
    \includegraphics[width=\linewidth]{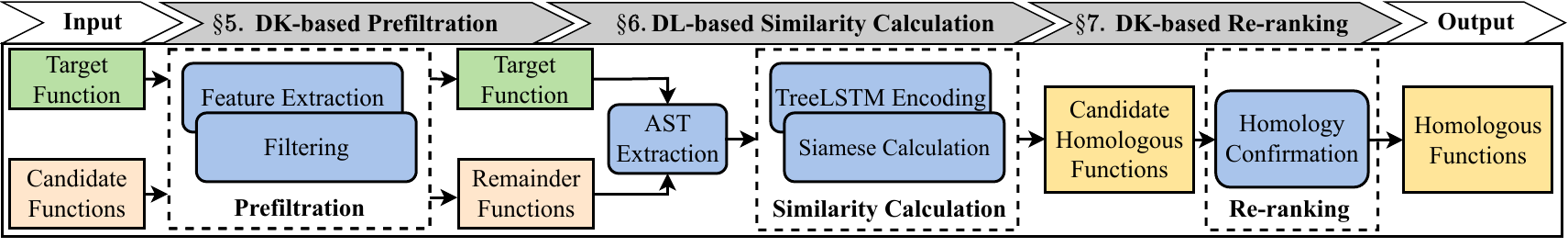}
    \caption{Workflow of \toolname{}. DK stands for Domain Knowledge. DL stands for Deep Learning.}
    \label{fig:workflow}
\end{figure}

\section{Methodology Overview}

\toolname{} consists of three primary modules: \textbf{DK-based Prefiltration}, \textbf{DL-based Similarity Calculation}, and \textbf{DK-based Re-ranking}, as shown in Figure~\ref{fig:workflow}. Here, DK stands for Domain Knowledge, and DL stands for Deep Learning.
The DK-based prefiltration module utilizes syntactic features to filter out dissimilar functions from the candidate functions in a lightweight and efficient manner (see \cref{sec:prefiltration}).
The DL-based similarity calculation module encodes ASTs into representation vectors using the Tree-LSTM model and determines the similarity score between the target function and the remaining functions using a Siamese network (see \cref{sec:func_similarity}).
The DK-based re-ranking module reorders the candidate homologous functions produced by the DL-based similarity calculation module using lightweight structural features, such as the function call relationship.
By integrating these three modules, \toolname{} efficiently and effectively detects homologous functions across architectures.

\section{DK-based Prefiltration} \label{sec:prefiltration}
At this stage, \toolname{} aims to incorporate an efficient and effective filter. To achieve this goal, we have summarized the challenges associated with fully exploiting the \textit{NCL} feature. Based on these challenges, we have developed a novel algorithm that overcomes these obstacles and enables us to construct the filter.

\subsection{Exploitation Challenges}
We have manually examined the false-negative cases where homologous functions were filtered out by both the \textit{NCL} and \textit{No. of callee} features. Through this examination, we have identified the challenges associated with appropriately exploiting the \textit{NCL} and \textit{No. of callee} features to address these false-negative cases.

\begin{itemize}
    \item \textbf{Exploitation Challenge 1 (EC1).} Decorated callee function name. A decorated callee function name is the result of a function name being decorated by the compiler using various techniques~\cite{function_decoration}. One such technique is name mangling, which is used by the C++ compiler to encode the function name with additional information about its parameters and return type to facilitate function overloading. The name of the function after decoration may differ from its original name in the source code and will be distinct across different architectures, particularly X86 and X64, due to their differing return types.
        
    \item \textbf{Exploitation Challenge 2 (EC2).} In certain functions, commonly referred to as leaf nodes in a call graph, there are no callee functions present. These functions are self-contained and do not call any other functions within their code. As a result, the leaf nodes do not possess distinguishable \textit{NCL} and \textit{No. of callee} features.
    
    \item \textbf{Exploitation Challenge 3 (EC3).} Function calls in binary target functions might not always be consistent with source code. Function calls may be added or deleted due to compiler optimization.
    The reasons for the function call change are function inline, intrinsic function replacement, instruction replacement for optimization, that behave differently in different architectures. These challenges are introduced in ~\cref{sec:func_call_opt}.
    
\end{itemize}

To overcome exploitation challenges, we improve feature \textit{ECL} and propose a novel algorithm \textbf{UpRelation}.

\subsection{Definition of \textit{NCL}}  \label{sec:new_feature}

This section provides a formal definition of \textit{NCL} to enhance clarity and precision. 
\textit{NCL} is built upon the call graph of the software.
The \textit{call graph} $CG$ can be defined by representing all functions as nodes and the call relationships between them as edges: $CG = (\mathcal{V},\mathcal{E})$, where $\mathcal{V} = \{v| \text{$v$ is a function}\}$ denotes the node collection and $\mathcal{E} = \{ (u,v)| \text{$u$ calls $v$}\}$ denotes the edge collection. For any edge $(u,v) \in \mathcal{E}$, we say that function $v$ is a callee function of function $u$.
To facilitate linking, function names in the dynamic symbol table $DST$ (i.e., import and export table) are preserved~\cite{harris2005practical}. For instance, if a target function calls an external function such as `strcpy', the callee function name `strcpy' remains in the import table, rather than being removed after binary stripping.
The \textit{NCL} of a target function $f$ is defined as $NCL_f = \{ v  | v \in \mathcal{V},  v \in DST \, (f, v) \in  \mathcal{E}, \}$, where $v$ is sorted by its call instruction address.

To address EC1, we employ two strategies to recover the original function names. Firstly, for C++ decorated names, we use the recovery tool \textit{cxxfilt}~\cite{cxxfilt} to recover the function names. Secondly, for other decorated functions, we define heuristic rules to recover the function names. For example, we recover the function call to '\_gets' by replacing it with 'gets', by removing the underscore at the beginning. In cases where a function calls the same function multiple times, we keep multiple identical function names.

\subsection{Filtration algorithm} \label{sec:filtermethod}

To address the additional two challenges, we propose a callee similarity-based algorithm called \textbf{UpRelation}. This algorithm leverages context information in the call graph to overcome challenges \textbf{EC2} and \textbf{EC3}.
Specifically, the algorithm utilizes parent nodes of leaf nodes in the call graph to match similar leaf nodes and address challenge \textbf{EC2}.
In the algorithm, we adopt a drill-down strategy that combines three features: \textit{NCL}, \textit{No. Callee}, and \textit{String Constants}, based on their information content.
The \textit{No. Callee} of function $f$ is denoted by $Callee_f$, and the set of \textit{String Constants} for function $f$ is denoted by $StrCons_f$.

\begin{algorithm}
\caption{UpRelation}
    \label{alg:uprelaiton}
    \SetKwFunction{RDR}{RDR} 
    \SetKwFunction{CSR}{CSR} 
    \SetKwFunction{GetCallers}{GetCallers}
    \SetKwFunction{GetCallees}{GetCallees}
    \KwIn{Vulnerable Function $fv$, Target Function List $TFL$, Thresholds $T_{NCL}, T_{callee}, T_{string}$}
    \KwOut{Vulnerable Candidate Function List $VFL$}
    
    $VFL \gets TFL$\;
    
    \uIf{$NCL_{fv}$ is not null}{
        \For{$f \in TFL $}
        {   
            $s$ = \CSR{$NCL_{fv}$, $NCL_{f}$}\;
            \lIf{$s < T_{NCL}$}{$VFL$.pop($f$)}
        }
    }
    \uElseIf{$ Callee_{fv} > 0$ }{
        \For{$f \in TFL $}{
            $s$ = \RDR{$Callee_{fv}$, $Callee_{f}$}\;
            \lIf{$s < T_{callee}$}{$VFL$.pop($f$)}
        }
    }
    \uElse{
        $FL'$ = $\emptyset$\;
        \For{$caller \in $ \GetCallers{fv}}{
            \For{$caller' \in UpRelation(caller, VFL)$}{
                $FL'$.add(\GetCallees{$caller'$})\;
            }
        }
        $VFL$ = $FL'$\;
        }
        
    \eIf{$StrCons_{fv}$ is not null}{
    \For{$f \in VFL$}{
         $s$ = \CSR{$StrCons_{fv}$, $StrCons_{f}$}\;
          \lIf{$s < T_{string}$}{$VFL$.pop($f$)}
      }    
  }
\Return $VFL$\;

\end{algorithm}

Given a vulnerable function $f_v$, Algorithm~\ref{alg:uprelaiton} aims to eliminate most non-homologous functions while retaining the vulnerable candidate functions in a list ($VFL$) from the target function list ($TFL$).
The code from lines 2 to 6 performs filtering when the feature $NCL_{fv}$ of $f_v$ is not empty. Specifically, the algorithm calculates the callee similarity ratio ($CSR$) between $NCL_{fv}$ and $NCL_f$ of all candidate functions in line 4. It then filters out functions whose $CSR$ is less than a threshold $T_{NCL}$.
Similarly, when the number of callee functions ($Callee_{fv}$) of $f_v$ is not zero, the algorithm filters out functions by calculating the Relevance Distance Ratio ($RDR$) score from line 7 to 11.
The most crucial portion of the algorithm is in lines 12 to 19, where it matches the leaf functions to address \textbf{EC2}. All caller functions of $f_v$ are first visited, and the algorithm employs $UpRelation$ to discover all functions that are similar to the caller function $caller$ in line 14. For each similar function $caller'$, the algorithm considers all its callee functions as vulnerable candidate functions at line 16.
Matching the same leaf functions by locating the same caller functions introduces some extraneous (leaf) functions that share the same caller function but are not the same as the leaf function. To remove these extraneous functions, the algorithm utilizes string similarity at line 22.
After filtering by callees and strings, the algorithm finally obtains the expected vulnerable candidate function list $VFL$.

\begin{figure}[h]
    \centering
    \includegraphics[width=0.6\linewidth]{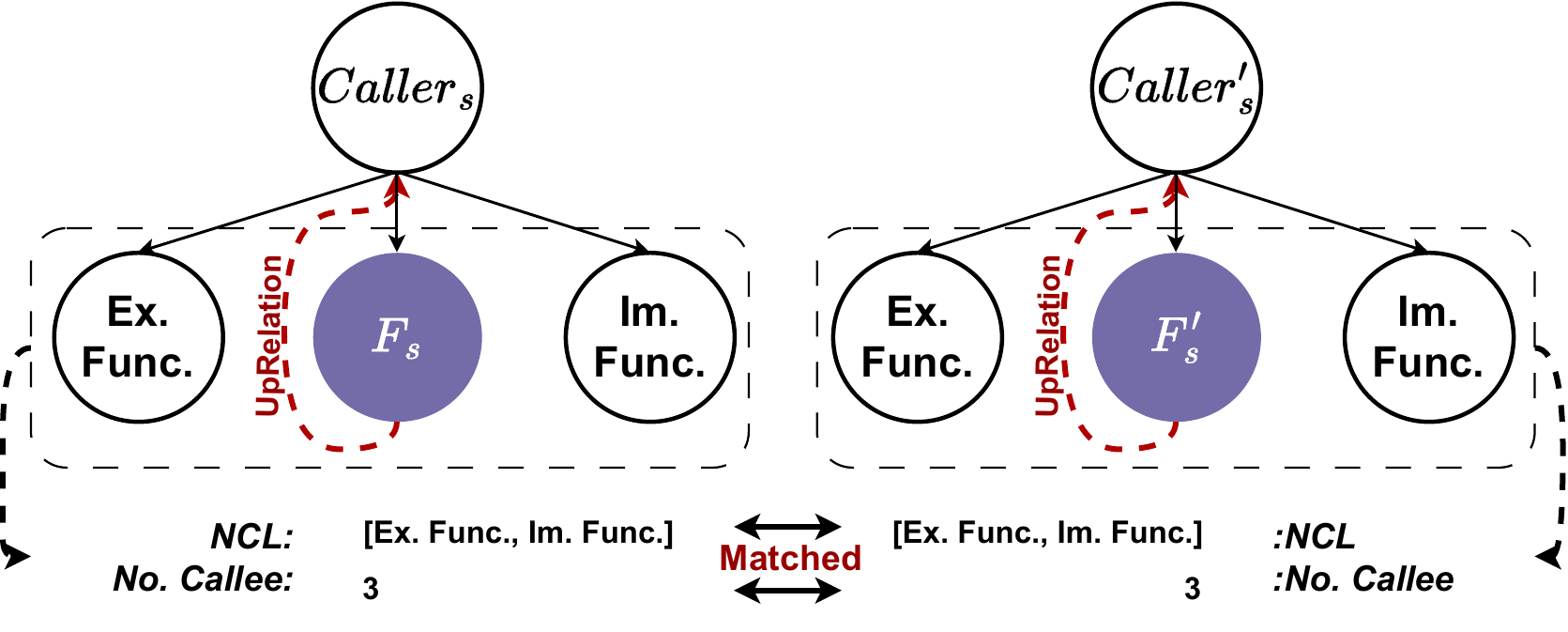}
    \caption{The Leaf Function $F_s'$ is Matched by Its Homologous Function $F_s$ (rather than being filtered out). \textbf{Ex. Func.} is short for `Exported Function'. \textbf{Im. Func.} is short for `Imported Function'. }\label{fig:uprelation}
\end{figure}
\textbf{Leaf Node Calculation Illustration.}
When the \textit{Strings}, \textit{No. Callee}, and \textit{NCL} are non-empty, the similarity calculation in our algorithm is straightforward. The arduous aspect of the algorithm lies in managing leaf functions that do not call other functions. To provide a clearer illustration, we have employed an example depicted in Figure~\ref{fig:uprelation} to demonstrate why homologous functions of leaf function $F_s$ are preserved after pre-filtration. In this example, we presume that leaf function $F_s$ does not comprise any strings. The algorithm proceeds to lookup its caller and collate its \textit{NCL} as $[Ex.,Func., Im.,Func.]$, where \textbf{Ex. Func.} is an abbreviation for 'Exported Function', and \textbf{Im. Func.} is an abbreviation for 'Imported Function'. Similarly, the algorithm collects the \textit{NCL} of caller of $F_s'$ and attempts to correlate between the two \textit{NCL}s. We postulate that homologous functions from the same software have equivalent callers, signifying that caller functions $caller$ and $caller'$ invoke the same exported and imported functions. Consequently, the \textit{NCL} of two caller functions comprise the same elements $[Ex.,Func., Im.,Func.]$. Upon the successful correlation of the \textit{NCL} of caller function $Caller_s'$, the algorithm preserves all its offspring nodes, encompassing $F_s'$, \textbf{Im. Func.}, and \textbf{Ex. Func.}, and eliminates all other functions. As a result, homologous function $F_s'$ of $F_s$ is conserved after pre-filteration.

\begin{figure}[h]
    \centering
    \includegraphics[width=0.7\linewidth]{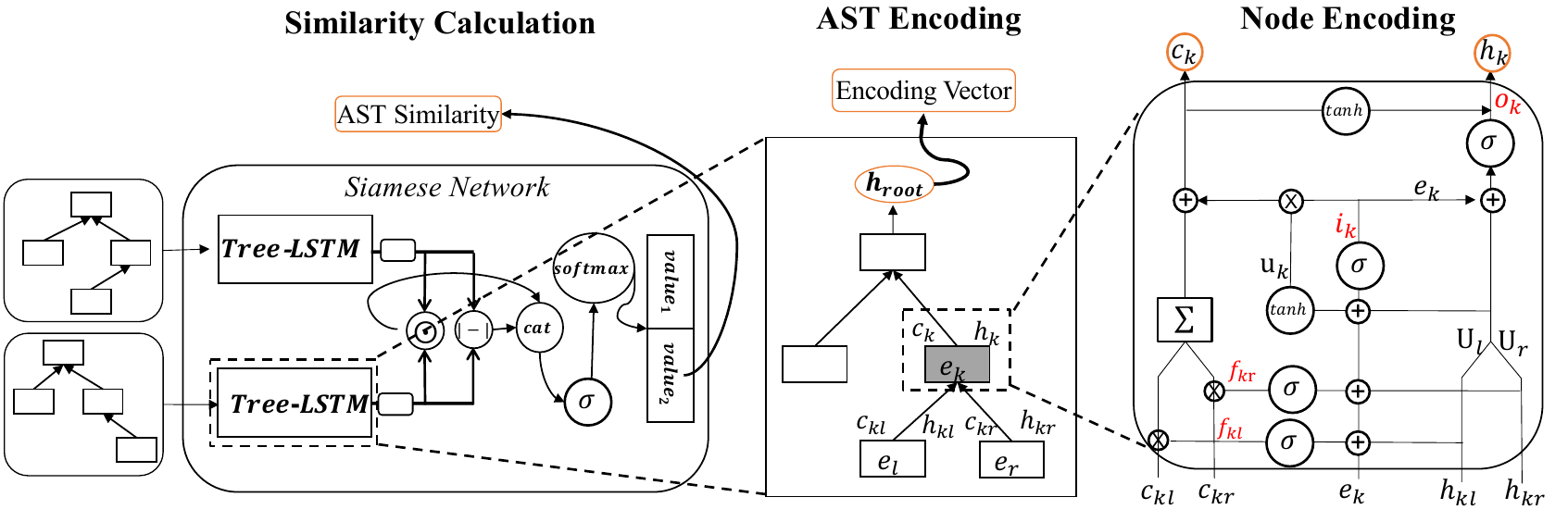}
    \caption{The Siamese Architecture and Tree-LSTM Encoding.}
    \label{fig:siamese}
\end{figure}

\section{DL-based Similarity Calculation} \label{sec:func_similarity}
This module calculates the similarity between two function ASTs by encoding them into vectors and applying the \siamese{} to calculate similarity between encoded vectors. Figure~\ref{fig:siamese} depicts the calculation flow.

\subsection{Tree-LSTM Encoding} \label{sec:treelstm}
Given an AST, Tree-LSTM model encodes it into a representation vector.
Tree-LSTM model is firstly proposed to encode the tree representation of a sentence and summarize the semantic information in natural language processing.
Tree-LSTM model can preserve every property of the plain LSTM gating mechanisms while processing tree-structured inputs.
The main difference between the plain LSTM and the Tree-LSTM is the way to deal with the outputs of predecessors.
The plain LSTM utilizes the output of only one predecessor in the sequence input.
We utilize Tree-LSTM to integrate the outputs of all child nodes in the AST for calculation of the current node. 
To facilitate the depiction of the Tree-LSTM encoding, we assume that node $v_k$ has two child nodes $v_l$ and $v_r$.
The  Tree-LSTM  encoding of node $v_k$ takes three types of inputs: node embedding $e_{k}$ of $v_k$, hidden states $h_{kl}$ and $h_{kr}$, and cell states $c_{kl}$ and $c_{kr}$ as illustrated in Figure~\ref{fig:siamese}.
The node embedding $e_{k}$ is generated by using the pre-trained model CodeT5 to embed the node $v_k$ to a high-dimensional representation vector.
$h_{kl}$, $h_{kr}$, $c_{kl}$, and $c_{kr}$ are outputs from the encoding of child nodes.
During the node encoding in Tree-LSTM, there are three gates and three states which are important in the calculation.
The three gates are calculated for filtering information to avoid gradient explosion and gradient vanishing~\cite{tai2015improved}.
They are input, output, and forget gates.
There are two forget gates $f_{kl}$ and $f_{kr}$, filtering the cell states from the left child node and right child node separately.
As shown in \textbf{Node Encoding} in Figure \ref{fig:siamese}, the forget gates are calculated by combining $h_{kl}$, $h_{kr}$, and $e_k$.
Similar to the forget gates, the input gate, and the output gate are also calculated by combining $h_{kl}$, $h_{kr}$, and $e_k$.
The details of the three types of gates are as follows:
\begin{equation}
\label{equation:fl}
\begin{aligned}
   f_{kl} = \sigma(W^{f}e_k+(U^{f}_{ll}h_{kl}+U^{f}_{lr}h_{kr})+b^{f})
\end{aligned}
\end{equation}
\begin{equation}
\label{equation:fr}
\begin{aligned}
   f_{kr} = \sigma(W^{f}e_k+(U^{f}_{rl}h_{kl}+U^{f}_{rr}h_{kr})+b^{f})
\end{aligned}
\end{equation}
\begin{equation}
\label{equation:ik}
\begin{aligned}
   i_{k} = \sigma(W^{i}e_k+(U^{i}_lh_{kl}+U^{i}_rh_{kr})+b^{i})
\end{aligned}
\end{equation}
\begin{equation}
\label{equation:ok}
\begin{aligned}
   o_{k} = \sigma(W^{o}e_k+(U^{o}_{l}h_{kl}+U^{o}_{r}h_{kr})+b^{o})
\end{aligned}
\end{equation}
where $i_k$ and $o_k$ denote the input gate and the output gate respectively,
and the symbol $\sigma$ denotes the \texttt{sigmoid} activation function.
The weight matrix $W$, $U$, and bias $b$ are different corresponding to different gates.
After the gates are calculated, there are three states $u_k$, $c_k$,  and $h_k$ in Tree-LSTM to store the intermediate encodings calculated based on inputs $h_{kl}$, $h_{kr}$, and $e_k$.
The cached state $u_k$ combines the information from the node embedding $e_k$ and the hidden states $h_{kl}$ and $h_{kr}$ {(Equation \ref{equation:uk}).}
And note that $u_k$ utilizes \texttt{tanh} as the activation function rather than $sigmoid$ for holding more information from the inputs.
The cell state $c_k$ combines the information from the cached state $u_k$ and the cell states $c_{kl}$ and $c_{kr}$ filtered by forget gates (Equation \ref{equation:ck}).
The hidden state $h_k$ is calculated by {combining the information} from cell state $c_k$ and the output gate $o_k$ (Equation \ref{equation:hk}).
The three states are computed as follows:
\begin{equation}
\label{equation:uk}
\begin{aligned}
   u_{k} = tanh(W^{u}e_k+(U^{u}_{l}h_{kl}+U^{u}_{r}h_{kr})+b^{u})
\end{aligned}
\end{equation}
\begin{equation}
\label{equation:ck}
\begin{aligned}
   c_{k} =  i_k \odot u_k +( c_{kl} \odot f_{kl} + c_{kr} \odot f_{kr})
\end{aligned}
\end{equation}
\begin{equation}
\label{equation:hk}
\begin{aligned}
   h_k = o_k \odot tanh(c_k)
\end{aligned}
\end{equation}
where the $\odot$ means Hadamard product~\cite{horn1990hadamard}.
After the hidden state and input state are calculated, the encoding of the current node $v_k$ is finished.
The states $c_k$ and $h_k$ will then be used for the encoding of $v_k$'s parent node.
During the AST encoding, Tree-LSTM encodes every node in the AST from bottom up as shown in \textbf{Tree-LSTM Encoding} in Figure~\ref{fig:siamese}.
After encoding all nodes in the AST, the hidden state of the root node is used as the encoding of the AST.

\subsection{Siamese Calculation} \label{sec:siamese}
This step uses Siamese architecture that integrates two identical Tree-LSTM model to calculate similarity between encoded vectors.
The details of the \siamese{} $\mathcal{M}({T}_1, {T}_2)$ are shown in Figure \ref{fig:siamese}.
The \siamese{} consists of two identical Tree-LSTM networks that share the same parameters.
In the process of similarity calculation, the \siamese{} first utilizes Tree-LSTM to encode ASTs into vectors.
We design the \siamese{} with subtraction and multiplication operations to capture the relationship between the two encoding vectors.
After the operations, the two resulting vectors are concatenated into a larger vector.
Then the resulting vector goes through a layer of \textit{softmax} function to generate a 2-dimensional vector.
The calculation is defined as:
\begin{equation}
\label{equation:classification}
\begin{aligned}
   \mathcal{M}({T}_1, {T}_2) = softmax(\sigma( cat(|\mathcal{N}({T}_1) - \mathcal{N}({T}_2)|, \mathcal{N}({T}_1) \odot \mathcal{N}({T}_2)) \times W )))
\end{aligned}
\end{equation}
where $W$ is a $2n \times 2$ matrix, the $\odot$ represents Hadamard product~\cite{horn1990hadamard}, $|\cdot|$ denotes the operation of making an absolute value, the function $cat(\cdot)$ denotes the operation of concatenating vectors.
The softmax function normalizes the vector into a probability distribution.
Since $W$ is a $2n \times 2$ weight matrix, the output of \siamese{} is a $2\times 1$ vector.
The format of output is $[\small{dissimilarity\text{ }score},\small{similarity\text{ }score}]$, where the first value represents the dissimilarity score and the second represents the similarity score.
During the model training, the input format of \siamese{} is $<{T}_1, {T}_2, label>$. 
In our work, the label vector $[1,0]$ means ${T}_1$ and ${T}_2$ are from non-homologous function pairs and the vector $[0,1]$ means homologous. 
The resulting vector and the label vector are used for model loss and gradient calculation.
During model inference, the second value in the output vector is taken as the similarity of the two ASTs, and the similarity of ASTs is used in re-ranking.

\begin{figure}[h]
    \centering
    \includegraphics[scale=0.8]{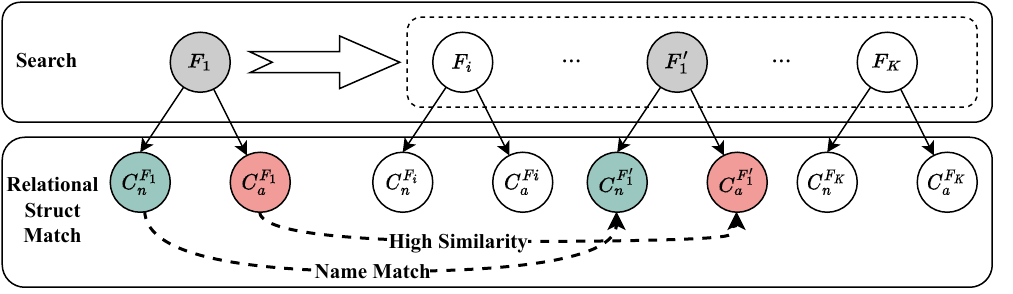}
    \caption{The Re-ranking Motivation Example. In the rectangular box with dashed line are the top K candidate homologous functions of $F_1$ produced by the search (i.e., DL-based similarity detection).
    Solid line arrows indicate the function call relationship (e.g., $F_1$ calls $C_{n}^{F_1}$). 
    The dotted line arrows indicate the callee function match in re-ranking.}
    \label{fig:rerank_overview}
\end{figure}

\section{DK-based Re-ranking}\label{sec:rerank}
This module seeks to confirm the homology of the top k candidate functions output by the Tree-LSTM network by re-ranking them.
In the prior phase, the Tree-LSTM network infers the semantic information from the AST, which is an intra-functional feature.
The knowledge gained from the AST is insufficient to establish the homology of functions. 
In this phase, function call relationships are used as domain knowledge to compensate for the lack of knowledge regarding the inter-functional features of the Tree-LSTM.
To this end, we design an algorithm called \textbf{Relational Structure Match}.
In contrast to the callee application in the pre-filtering module, this module uses more extensive information from callee relationships to show the degree of homology of candidate functions.

\subsection{Motivated Example}\label{sec:rerank_example}
Our algorithm is based on a conforming observation to an intuitive law: \textit{If a function $F_1$ calls function $C^{F_1}$, then its homologous function $F_1'$ will also call the homologous function $C^{F_1'}$ of $C^{F_1}$.}
As depicted in Figure~\ref{fig:rerank_overview}, we have $F_1$ calls $C^{F_1}$, and $F_1'$ calls $C^{F_1'}$.
Assume that the search process for $F_1$ yields the top K functions containing the target homologous function $F_1'$.
We then employ the call relations of $F_1$ and $F_1'$ to conduct precise callee function matching for re-ranking.
In particular, callee functions of $F_1$ are divided into two categories, named callees $C_{n}^{F_1}$ and anonymous callees $C_{a}^{F_1}$.
For named callees, their names are utilized to match callees of functions between the source function $F_1$ and the candidate top K functions.
For anonymous callees, we employ DL-based similarity detection to calculate the similarity between callees of functions between the source function $F_1$ and the candidate top K functions.
Recalling the observation, the homologous function $F_1'$ of $F_1$ holds the most matched callees.
After re-ranking the candidate functions based on the matched callees, $F_1'$ is re-ranked in the first place.

\subsection{Relational Structure Match Algorithm} \label{sec:relation_structure_match_algo}
The algorithm aims to rescore each candidate function by leveraging the call relationship between the target function and the candidate functions. 
The relational structure refers to the call relations between the target function and all its callee functions, as illustrated in Section~\cref{sec:rerank_example}.
To match the relational structure, the algorithm performs one of two distinct operations ($O_1$ and $O_2$) based on whether the source function has callee functions or not.
\begin{itemize}
    \item [\textbf{$O_1$}:] When the source function $F_1$ has one or more callee functions, the algorithm extracts all callee functions of $F_1$ to build a \textit{mixed callee function set} (\textit{MCFS}). (details are described below). Using \textit{MCFS}, the algorithm calculates similarities between the target function and the candidate functions, resulting in new match scores. 
    It re-ranks all candidate functions by combining the original \textbf{Asteria} scores (Equation~\ref{eq:fscore}) with the newly calculated match scores. 
    The details of \textit{MCFS} and match score calculation are described in the subsequent sections.
    \item [\textbf{$O_2$}:] When the source function $F_1$ has no callee functions, the algorithm removes all candidate functions that have one or more callee functions. The remaining candidate functions are then re-ranked based on their original \textbf{Asteria} scores.
\end{itemize}

\subsubsection{Mixed Callee Function Set.}
The \textit{mixed callee function set} (\textit{MCFS}) of function $F$ consists of two types of callee functions: named callee functions and anonymous callee functions.
Named callee functions refer to functions whose names have been preserved. These functions are typically imported or exported functions, and their function names are necessary for external linking purposes.
On the other hand, anonymous callee functions are a type of function for which the function name has been removed for security reasons. These functions are typically anonymized to protect sensitive information.
We denote the \textit{MCFS} of function $F$ as $CS_F = {C_{n1}^{F}, ...,C_{nj}^{F}, C_{a1}^{F}, ..., C_{aj}^{F}}$, where $C_{nj}^{F}$ represents a named callee function and $C_{aj}^{F}$ represents an anonymous callee function. The set $CS_F$ includes both types of callee functions for function $F$.


\subsubsection{Match Score Calculation.} 




The algorithm performs two types of matches to calculate the match score $M$ for each candidate function, utilizing the \textit{MCFS}s of the target function $F_1$ and all candidate functions.

\textbf{Named Callee Match:} For all named callees $C_{nj}^{F_1}$ in $CS_{F_1}$, the algorithm matches them with the named callees of each candidate function based on function names. If a named callee $C_n^{F_1}$ in $F_1$ has the same function name as a named callee $C_n^{F_1'}$ in a candidate function $F_1'$, they are considered a match. The number of matched functions in candidate function $F_i$ is denoted as $\mathcal{N}_n^{F_i}$.

\textbf{Anonymous Callee Match:} For all anonymous callees $C_{ai}^{F_1}$ in $CS_{F_1}$, the algorithm utilizes DL-based similarity detection to calculate similarity scores between all anonymous callees of the target function and the anonymous callees of all candidate functions. For each anonymous callee $C_{aj}^{F_i}$ in a candidate function $F_i$, the algorithm calculates the maximum similarity score between $C_{aj}^{F_i}$ and all anonymous callees of the target function. This maximum similarity score is denoted as $\mathcal{S}_{aj}^{F_i}$.

After matching all callee functions of the candidate functions, the match score $M_{F_i}$ of candidate function $F_i$ is calculated as follows:
\begin{equation}\label{eq:match}
M_{F_i} = \mathcal{N}_n^{F_i} + \sum{\mathcal{S}_{aj}^{F_i}}
\end{equation}
where $\mathcal{S}_{aj}^{F_i}$ represents the similarity score between an anonymous callee $C_{aj}^{F_i}$ in candidate function $F_i$ and the anonymous callees of the target function $F_i$. The sum is taken over all anonymous callees in $CS_{F_i}$, the \textit{MCFS} of candidate function $F_i$.

\subsubsection{Match Score-based Re-ranking.}

The re-ranking score of candidate function $F_i$ is obtained by combining the match score $M_{F_i}$ and its DL-based similarity score $\mathcal{M}{F_i}$ using Equation~\ref{eq:rerank}. The algorithm calculates a new score $S^{re-rank}{F_i}$ for each candidate function $F_i$ as follows:
\begin{equation}\label{eq:rerank}
    S^{re-rank}_{F_i} = \alpha \times \mathcal{M_{F_i}} + \beta \times M_{F_i}
\end{equation}
Here, $\alpha$ and $\beta$ are weight coefficients that satisfy $\alpha + \beta = 1$. 
The new score $S^{re-rank}{F_i}$ combines the DL-based similarity score $\mathcal{M}{F_i}$ and the match score $M_{F_i}$, emphasizing their importance according to the weights. 
A higher re-ranking score indicates a higher degree of homology.

After calculating the re-ranking scores for all candidate functions, the algorithm sorts them in descending order based on their new scores $S^{re-rank}_{F_i}$. This ranking allows for the identification of candidate functions with higher homology, as those with higher scores are prioritized.

\section{Evaluation}
We aim to conduct a comprehensive practicality evaluation of various state-of-the-art function similarity detection methods for bug search.
To this end, we adopt 8 different metrics to depict the search capability of different methods in a more comprehensive way.
Furthermore, we construct a large evaluation dataset, in a way that is closer to practical usage of bug search.

\subsection{Research Questions}
In the evaluation experiments, we aim to answer following research questions:
\begin{itemize}
    \item [\textbf{RQ1.}] How does \toolname{} compare to baseline methods in cross-architecture and cross-compiler function similarity detection?
    \item [\textbf{RQ2.}] What is the performance of \toolname{}, compared to baseline methods for bug search purpose?
    \item [\textbf{RQ3.}] How much do DK-based filtration and DK-based re-ranking improves in accuracy and efficiency for \toolname{}? How do their performance compare to other baseline methods when integrated together?
    \item[\textbf{RQ4.}]  How do different configurable parameters affect the accuracy and efficiency?
    \item [\textbf{RQ5.}] How does \toolname{} perform in a real-world bug search?
\end{itemize}

\subsection{Implementation Details}

We utilize \texttt{IDA Pro 7.5} \cite{ida} and its plugin \textit{Hexray Decompiler} to decompile binary code and extract ASTs. The current version of the \textit{Hexray Decompiler} supports x86, x64, PowerPC (PPC), and ARM architectures.
For the encoding of leaf nodes in Formulas (\ref{equation:fl})-(\ref{equation:ck}), we assign zero vectors to the state vectors $h_{kl}$, $h_{kr}$, $c_{kl}$, and $c_{kr}$.
During model training, we use the binary cross-entropy loss function (\textit{BCELoss}) to measure the discrepancy between the labels and the predictions. The \textit{AdaGrad} optimizer is utilized for gradient computation and weight-matrix updating after the losses are computed.
Due to the dependency of Tree-LSTM computation steps on the AST shape, parallel batch computation is not possible. Therefore, the batch size is always set to 1.
The model is trained for 60 epochs.
Our experiments are conducted on a local server with two Intel(R) Xeon(R) CPUs E5-2620 v4 @ 2.10GHz, each with 16 cores, 128GB of RAM, and 4TB of storage. The \toolname{} code runs in a Python 3.6 environment. We compile the source code in our dataset using the \texttt{gcc v5.4.0} compiler and utilize \texttt{buildroot-2018.11.1}~\cite{url:buildroot} for dataset construction. We use the binwalk tool~\cite{url:binwalk} to unpack firmware and obtain the binaries for further analysis.
In the \textbf{UpRelation} algorithm of the filtering module, we set the threshold values $T_{NCL}, T_{callee}, T_{string}$ to 0.1, 0.8, and 0.8, respectively, based on their $F_{score}$. The crucial threshold $T_{NCL}$ is discussed in \cref{sec:threshold_eval}.
In Equation~\ref{eq:rerank}, we set $\alpha = 0.1$ and $\beta = 0.9$ to emphasize the role of callee function similarities in the re-ranking process. The sensitivity analysis of these weights is presented in \cref{sec:weights_eva}.

\subsection{Comprehensive Benchmark}
To compare BCSD methods in a comprehensive way, we build an extensive benchmark based on multiple advanced works~\cite{xu2017neural, marcelli2022machine, wang2022jtrans}.
The benchmark comprises of two datasets, two detection tasks, and five measure metrics.

\subsubsection{Dataset}

The functions not involved in the prefiltering test (see \cref{sec:preliminary_study}) are divided into two datasets for model training and testing and evaluation.
The evaluation dataset consists of two sub-datasets, each of which is used for a different detection task.

\textbf{Model Dataset Construction.} The model dataset is constructed for training and testing the Tree-LSTM model. It consists of a total of 31,940 functions extracted from 1,944 distinct binaries. From these functions, 314,852 pairs of homologous functions and 314,852 pairs of non-homologous functions are created.
To ensure a fair evaluation of the model's performance, the dataset is divided into a training set and a testing set using an 8:2 ratio. This means that 80\% of the function pairs are used for training the model, while the remaining 20\% are used for testing and evaluating the model's performance.
The dataset construction allows the Tree-LSTM model to learn and generalize from a diverse set of functions, including both homologous and non-homologous pairs. By dividing the dataset into training and testing sets, the model's performance can be assessed on unseen data to measure its effectiveness in identifying homologous functions.

\textbf{Evaluation Dataset Construction.}
The dataset construction process involves creating two sub-datasets: the g-dataset and the v-dataset. These datasets are used for different evaluation tasks: classification test and bug search test.
The \textbf{g-dataset} is constructed for the classification test, which evaluates the model's ability to classify homologous and non-homologous function pairs. 
It consists of tuples in the form $(F, (F_h, F_n))$, where $F$ is the source function and $(F_h, F_n)$ represents a function set containing a homologous function $F_h$ and a non-homologous function $F_n$. 
Each tuple in the g-dataset represents a pair of functions to be classified as homologous or non-homologous.
On the other hand, the \textbf{v-dataset} is constructed for the bug search test, which evaluates the model's ability to identify non-homologous functions among a larger set of candidates. The tuples in the v-dataset are of the form $(F, (F_h, F_{n1}, ..., F_{ni}, ..., F_{n10000}))$. 
Here, $F_h$ represents a homologous function, and $F_{n1}$ to $F_{n10000}$ represent non-homologous functions.
In this case, the $P_{set}$ contains a larger number of non-homologous functions to simulate the bug search scenario.
For both datasets, the source function $F$ is matched with all the functions in the $P_{set}$ for evaluation. The g-dataset focuses on evaluating the model's accuracy in classifying homologous and non-homologous pairs, while the v-dataset assesses the model's performance in identifying non-homologous functions among a larger pool of candidates.

\subsubsection{Metrics}
We choose five distinct metrics for comprehensive evaluation from earlier works~\cite{yang2021asteria,wang2022jtrans,trex}.
In our evaluation, the similarity of a function pair is calculated as a score of $r$. 
Assuming the threshold is $\beta$, if the similarity score $r$ of a function pair is greater than or equal to $\beta$, the function pair is regarded as a positive result, otherwise a negative result.
For a homologous pair, if its similarity score $r$ is greater than or equal to $\beta$, it is a true positive \ysgbf{(TP)}. If a similarity score of $r$ is less than $\beta$, the calculation result is a false negative \ysgbf{(FN)}. 
For a non-homologous pair, if a similarity score $r$ is greater than or equal to $\beta$, it is a false positive \ysgbf{(FP)}.
When the similarity score $r$ is less than $\beta$, it is a true negative \ysgbf{(TN)}. 
These metrics are described as following:
\begin{itemize}
    \item \textbf{TPR.} TPR is short for true positive rate. TPR shows the accuracy of homologous function detection at threshold $\beta$. It is calculated as $TPR = \frac{TP}{TP+FN} $.
    \item \textbf{FPR.} FPR is short for false positive rate. FPR shows the accuracy of non-homologous function detection at threshold $\beta$. It is calculated as $FPR = \frac{FP}{FP+TN}$.
    \item \textbf{AUC.} AUC is short for area under the curve, where the curve is termed Receiver Operating Characteristic \ysgbf{(ROC)} curve. The ROC curve illustrates the detection capacity of both homologous and non-homologous functions as its discrimination threshold $\beta$ is varied. 
    AUC is a quantitative representation of ROC.
    
    \item \textbf{MRR.} MRR is short for mean reciprocal rank, which is a statistic measure for evaluating the results of a sample of queries, ordered by probability of correctness. It is commonly used in retrieval experiments. In our bug retrieval-manner evaluation, it is calculated as $MRR = \frac{1}{|P_{set}|}\sum_{F_{hi}\in P_{set}}\frac{1}{Rank_{F_{hi}}}$, where $Rank_{F_{hi}}$ denotes the rank of  function  $F_{hi}$ in pairing candidate set $P_{set}$, and $|P_{set}|$ denotes the size of $P_{set}$.
    \item \textbf{Recall@Top-k.} It shows the capacity of homologous function retrieve at top k detection results. The top k results are regarded as homologous functions (positive). It is calculated as follows:
    $$
     g(x) = \begin{cases}
   1 &\text{if } x = True \\
   0 &\text{if } x = False
    \end{cases} 
    $$
    $$
     Recall@k = \frac{1}{|F|}\sum g(Rank_{f_i^{gt}} \le k)
    $$
    To demonstrate the reliability of the ranking results, we adopt Recall@Top-1 and Recall@Top-10.

\end{itemize}

\subsubsection{Detection Tasks}
The two function similarity detection tasks based on BCSD applications are as follows:

Task-C (Classification Task): This task focuses on evaluating the ability of methods to classify function pairs as either homologous or non-homologous. It involves performing binary classification on the \textbf{g-dataset}, which contains tuples of the form $(F, (F_h, F_n))$, where $F_h$ represents a homologous function and $F_n$ represents a non-homologous function. The task evaluates the performance using three metrics: TPR, FPR, and AUC of the ROC curve. TPR and FPR are commonly used to measure the performance of binary classification models, while AUC provides an overall measure of the model's discriminative ability.

Task-V (Bug/Vulnerability Search Task): This task focuses on evaluating the ability of methods to identify homologous functions from a large pool of candidate functions. It uses the \textbf{v-dataset}, which contains tuples of the form $(F, (F_h, F_{n1}, ..., F_{ni}, ..., F_{n10000}))$, where $F_h$ represents a homologous function and $F_{ni}$ represents non-homologous functions. The task involves calculating function similarity between a source function $F$ and all functions in the $P_{set}$. The functions in $P_{set}$ can then be sorted based on similarity scores. The task evaluates the performance using three metrics: MRR, Recall@Top-1, and Recall@Top-10. MRR measures the rank of the first correctly identified homologous function, while Recall@Top-1 and Recall@Top-10 measure the proportion of cases where the correct homologous function is included in the top-1 and top-10 rankings, respectively.

These tasks provide a comprehensive evaluation of the methods' performance in distinguishing between homologous and non-homologous functions and identifying homologous functions from a large pool of candidates.

\subsection{Baseline Methods.}
We choose various representative cross-architectural BCSD works, that make use of AST or are built around deep learning encoding.
These BCSD works consist of \diaphora{}~\cite{diaphora}, \gemini{}~\cite{xu2017neural}, \methodname{SAFE}~\cite{massarelli2019safe}, and \methodname{Trex}~\cite{trex}.
Moreover, we also use our previous conference work \methodname{Asteria} as one of baseline methods.
We go over these works in more details below.


\paragraph*{\textbf{Diaphora}}
\diaphora{} performs similarity detection also based on AST.
\diaphora{} maps nodes in an AST to primes and calculates the product of all prime numbers.
Then it utilizes a \textit{difference function} to calculate the similarity between the prime products.
We download the \diaphora{} source code from github \cite{diaphora}, and extract \diaphora{}'s core algorithm for AST similarity calculation for comparison.
Noting that it would take a significant amount of time (several minutes) to compute a pair of functions with extremely dissimilar ASTs, we add a filtering computation before the prime difference.
The filtering calculates the AST size difference and eliminates function pairs with a significant size difference. We publish the improved Diaphora source code on our website~\cite{asteria_pro_github}.

\paragraph*{\textbf{Gemini}}
\gemini{} encodes ACFGs (attributed CFGs) into vectors with a graph embedding neural network.
The ACFG is a graph structure where each node is a vector corresponding to a basic block.
We have obtained \gemini{}'s source code and its training dataset.
Notice that in~\cite{xu2017neural} authors mentioned it can be retrained for a specific task, such as the bug search.
To obtain the best accuracy of \gemini{}, we first use the given training dataset to train the model to achieve the best performance.
Then we re-train the model with the part of our training dataset.
\gemini{} supports similarity detection on X86, MIPS, and ARM architectures.

\paragraph*{\textbf{SAFE}}
\methodname{SAFE} works directly on disassembled binary functions, does not require manual feature extraction, is computationally more efficient than \gemini{}. In their vulnerability search task, \methodname{SAFE} outperforms \gemini{} in terms of recall.
\methodname{SAFE} supports three different instruction set architecture X64, X86, and ARM.
We retrain \methodname{SAFE} based on the official code~\cite{massarelli2019safe} and use retrained model parameter for our test.
In particular, we select all appropriate function pairs from the training dataset, whose instruction set architectures are supported by \methodname{SAFE}.
Then we extract the function features for all function pairs selected and discard the function pairs whose features \methodname{SAFE} cannot extract.
After feature extraction, 27,580 function pairs of three distinct architecture combinations (i.e., X86-X64, X86-ARM, and X64-ARM) are obtained for training.
Next, We adopt the default model parameters (e.g., embedding size) and training setting (e.g. training epoches) to train \methodname{SAFE}.

\paragraph*{\textbf{Trex}}
\methodname{Trex} is based on pretrained model~\cite{trex} of the state-of-the-art NLP technique, and micro-traces.
It utilizes a dynamic component to extract micro-traces and use them to pretrain a masked language model.
Then it integrates pretrained ML model into a similarity detection model along with the learned semantic knowledge from micro-traces.
It supports similarity detection of ARM, MIPS, X86, and X64.

\subsection{Comparison of Similarity Detection Accuracy (RQ1)} \label{sec:rq1_eval}
In the evaluation of cross-architecture scenarios, the focus was on assessing the detection capability of different approaches in two tasks separately, which is commonly encountered in vulnerability search scenarios. Additionally, the evaluation also considered the performance in cross-compiler scenarios involving three different combinations of compilers: gcc-clang, gcc-icc, and clang-icc.

\subsubsection{Cross-Architecture Evaluation}
In the evaluation of the two distinct tasks, it is important to note that the baseline methods may not be capable of detecting function similarities for all four instruction set architectures. As a result, the detection results for certain architecture combinations may be empty, indicating that the baseline methods were unable to provide any meaningful results.
For each task, the evaluation measured the performance of various approaches in terms of the defined metrics. The specific outcomes and results of the evaluation for each task were analyzed and discussed. 

\begin{figure}
    \centering
    \begin{minipage}[c]{0.3\linewidth}
    \includegraphics[width=\linewidth,page=1]{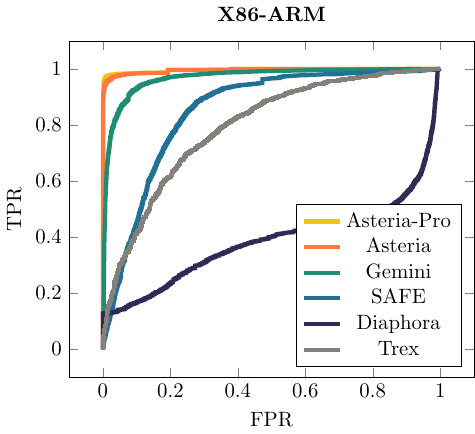}
    \end{minipage}
     \begin{minipage}[c]{0.3\linewidth}
    \resizebox{\linewidth}{!}{
    \includegraphics[width=\linewidth,page=2]{pics/rocs.pdf}
    }
    \end{minipage}
    \begin{minipage}[c]{0.3\linewidth}
    \resizebox{\linewidth}{!}{
    \includegraphics[width=\linewidth,page=4]{pics/rocs.pdf}
    }
    \end{minipage}
    \begin{minipage}[c]{0.3\linewidth}
    \resizebox{\linewidth}{!}{
    \includegraphics[width=\linewidth,page=5]{pics/rocs.pdf}
    }
    \end{minipage}
    \begin{minipage}[c]{0.3\linewidth}
    \resizebox{\linewidth}{!}{
    \includegraphics[width=\linewidth,page=7]{pics/rocs.pdf}
    }
    \end{minipage}
    \begin{minipage}[c]{0.3\linewidth}
    \resizebox{\linewidth}{!}{
    \includegraphics[width=\linewidth,page=8]{pics/rocs.pdf}
    }
    \end{minipage}
    \caption{ROC Curves on All Cross-architecture Combination Detection.}
    \label{fig:task-c-roc}
\end{figure}

\begin{table}[t]
\caption{AUCs in Task-C.} \label{tab:auc}
\begin{tabular}{c|lllllll}
\hline
\multicolumn{1}{l|}{\textbf{Methods}} & \textbf{X86-ARM} & \textbf{X86-X64} & \textbf{X86-PPC} & \textbf{ARM-X64} & \textbf{ARM-PPC} & \textbf{X64-PPC} & \textbf{Average} \\ \hline
\textbf{\toolname{}}                                           & \textbf{0.996}   & \textbf{0.998}   & \textbf{0.995}   & \textbf{0.998}   & \textbf{0.998}   & \textbf{0.999}   & \textbf{0.997}   \\ 
Asteria                                                        & {0.995}   & {0.998}   & {0.998}   & {0.995}   & {0.998}   & {0.999}   & {0.997}   \\ 
Gemini                                                         & 0.969            & 0.984            & 0.984            & 0.973            & 0.968            & 0.984            & 0.977            \\ 
SAFE                                                           & 0.851            & 0.867            & -                & 0.872            & -                & -                & 0.863            \\ 
Trex                                                           & 0.794            & 0.891            & -                & 0.861            & -                & -                & 0.849            \\ 
Diaphora                                                       & 0.389            & 0.461            & 0.397            & 0.388            & 0.455            & 0.400            & 0.415            \\ \hline
\end{tabular}
\end{table}

\paragraph*{\textbf{Comparison on task-C}}

In Task-C, all approaches were evaluated by conducting similarity detection on all supported architectural combinations. The evaluation results were used to calculate the three metrics (TPR, FPR, and AUC) for each approach. These results are presented in Table~\ref{tab:auc} and visualized in Figure~\ref{fig:task-c-roc}, where each subplot represents the ROC curve for a specific architecture combination. The x-axis represents the FPR (False Positive Rate), and the y-axis represents the TPR (True Positive Rate).
By examining the ROC curves in Figure~\ref{fig:task-c-roc}, it can be observed that methods with performance curves closer to the upper-left corner generally exhibit superior performance. In particular, the ROC curves of \toolname{} and Asteria are almost indistinguishable across all architectural combinations, indicating that they possess equivalent classification performance in Task-C.
Furthermore, the AUC values presented in Table~\ref{tab:auc} provide a quantitative measure of the approaches' ability to distinguish between homologous and non-homologous functions. It is noted that Asteria-Pro and Asteria demonstrate nearly identical performance in this regard. However, the AUC values of \toolname{} are consistently greater than those of the other baseline techniques for all architectural combinations. This suggests that \toolname{} exhibits superior discriminative capability between homologous and non-homologous functions in Task-C.
These findings highlight the strong performance of \toolname{} in the classification task and its ability to outperform the baseline methods in distinguishing between homologous and non-homologous functions across various architecture combinations.


\begin{table}[h]
\caption{MRR and Recall of Different Methods.}\label{tab:mrr_recall}
\resizebox{!}{3cm}{
\begin{tabular}{c|llllllll}
\hline

\multicolumn{1}{c|}{\textbf{Metrics}} & \textbf{Methods} & \textbf{X86-X64} & \textbf{X86-ARM} & \textbf{X86-PPC} & \textbf{X64-ARM} & \textbf{X64-PPC} & \textbf{ARM-PPC} & \textbf{Avg}   \\ \hline
                                                               & \textbf{\toolname{}}  & \textbf{0.934} & \textbf{0.887} & \textbf{0.931} & \textbf{0.879} & \textbf{0.919} & \textbf{0.903} & {\textbf{0.908}} \\ 
                                                               & Asteria          & 0.776   & 0.724   & 0.731   & 0.708   & 0.713   & 0.750   & {0.734} \\ 
                                                               & Trex             & 0.414   & 0.206   & -       & 0.309   & -       & -       & {0.310} \\ 
                                                               & Gemini           & 0.478   & 0.250   & 0.325   & 0.336   & 0.357   & 0.256   & {0.334} \\ 
                                                               & Safe             & 0.029   & 0.007   & -       & 0.009   & -       & -       & 0.015   \\ 
\multirow{-6}{*}{MRR}                                          & Diaphora         & 0.023   & 0.019   & 0.020   & 0.019   & 0.020   & 0.021   & {0.020} \\ \hline
                                                               & \textbf{\toolname{}}  & \textbf{0.917} & \textbf{0.868} & \textbf{0.912} & \textbf{0.879} & \textbf{0.899} & \textbf{0.903} & {\textbf{0.896}} \\ 
                                                               & Asteria          & 0.706   & 0.648   & 0.652   & 0.627   & 0.631   & 0.675   & {0.657} \\ 
                                                               & Trex             & 0.274   & 0.110   & -       & 0.192   & -       & -       & {0.192} \\ 
                                                               & Gemini           & 0.405   & 0.180   & 0.242   & 0.261   & 0.279   & 0.229   & {0.266} \\ 
                                                               & Safe             & 0.004   & 0.002   & -       & 0.002   & -       & -       & {0.003} \\ 
\multirow{-6}{*}{Recall@Top-1}                                     & Diaphora         & 0.021   & 0.016   & 0.017   & 0.016   & 0.017   & 0.018   & {0.018} \\ \hline
                                                               & \textbf{\toolname{}} & \textbf{0.961} & \textbf{0.921} & \textbf{0.962} & \textbf{0.913} & \textbf{0.952} & \textbf{0.932} & {\textbf{0.940}} \\  
                                                               & Asteria          & 0.902   & 0.867   & 0.882   & 0.857   & 0.866   & 0.890   & {0.877} \\ 
                                                               & Trex             & 0.710   & 0.452   & -       & 0.575   & -       & -       & {0.579} \\ 
                                                               & Gemini           & 0.615   & 0.383   & 0.482   & 0.478   & 0.502   & 0.468   & {0.488} \\  
                                                               & Safe             & 0.022   & 0.010   & -       & 0.014   & -       & -       & {0.015} \\  
\multirow{-6}{*}{Recall@Top-10}                                    & Diaphora         & 0.029   & 0.024   & 0.026   & 0.026   & 0.025   & 0.027   & {0.026} \\ \hline
\end{tabular}
}
\end{table}

\paragraph*{\textbf{Comparison on task-V}}

Table~\ref{tab:mrr_recall} presents the results of calculating MRR, Recall@Top-1, and Recall@Top-10 for different architectural combinations. These metrics evaluate the performance of the methods in the bug (vulnerability) search task. Recall@Top-1 measures the ability to accurately detect homologous functions, while Recall@Top-10 assesses the capability to rank homologous functions within the top ten positions.
In the table, the first column represents the metrics, and the second column lists the names of the methods. The third through eighth columns display the metric values for the different architectural combinations, while the last column shows the mean value across all architectures.
It can be observed that \toolname{} and \methodname{Asteria} consistently outperform the baseline approaches by a significant margin across all architecture configurations. 
\toolname{} achieves an impressive average MRR of 0.908, indicating a substantial improvement of up to 23.71\% compared to \methodname{Asteria}. 
Even after retraining, \methodname{Safe} demonstrates poor performance in properly recognizing small functions.
In terms of Recall@Top-1, both \toolname{} and \methodname{Asteria} achieve relatively high average precisions of 0.89 and 0.65, respectively, which are 237\% and 146\% higher than the best result (0.26). Notably, \toolname{} shows a 36.4\% improvement in Recall@Top-1 compared to \methodname{Asteria}.
Regarding Recall@Top-10, both \toolname{} and \methodname{Asteria} continue to exhibit superior performance compared to the other methods. While other methods show a significant increase in recall compared to Recall@Top-1, their values remain below \toolname{}.
Overall, these results demonstrate that \toolname{} outperforms the baseline methods, including \methodname{Asteria}, in terms of MRR, Recall@Top-1, and Recall@Top-10 across different architecture combinations. The recall of other methods, such as \methodname{Trex}, increases significantly from Recall@Top-1 to Recall@Top-10, indicating their ability to rank homologous sequences more accurately. However, they still fall short compared to \toolname{}.

Indeed, the performance of BCSD approaches can vary significantly between different evaluation tasks, as demonstrated by the differences in Task-V performance compared to the similar ROC curve performance in Task-C. In the case of \methodname{Gemini}, despite having a high AUC score similar to \methodname{Asteria}, its MRR performance is relatively poor compared to both \methodname{Asteria} and \toolname{}. This indicates that evaluating BCSD approaches in a single experiment setting, such as Task-C, may not provide a comprehensive understanding of their real-world applicability and behavior.
Task-V, which focuses on bug (vulnerability) search, simulates the scenario of identifying homologous functions from a pool of candidate functions. In this task, the ability to accurately rank and identify homologous functions becomes crucial. While ROC curves and AUC scores provide information about the ability to discriminate between homologous and non-homologous functions, they may not reflect the performance in ranking and retrieving homologous functions accurately.
Therefore, it is important to consider multiple evaluation tasks, such as Task-C and Task-V, to assess the overall performance and effectiveness of BCSD approaches. The results obtained from different tasks can provide a more comprehensive understanding of the strengths and limitations of each method and their suitability for real-world applications.

\begin{figure}
    \begin{minipage}{0.4\textwidth}
\begin{lstlisting}[language=C, escapechar=@]
CK_RV proxy_C_@\textcolor{red}{\textbf{DigestInit}}@(...){
    /*
    Variable Initialization.
    */
    v5 = (Proxy *)self[1].C_GetSlotInfo;
    v7 = handle;
    result = map_session_to_real(v5,&v7,&map,V3);
    if (!result)
        result = map.funcs->C_@\textcolor{red}{\textbf{DigestInit}}@(v7, mechanism);
    return result;
}
\end{lstlisting}
    \end{minipage}
    \hspace{0.5cm}
      \begin{minipage}{0.4\textwidth}
\begin{lstlisting}[language=C, escapechar=@]
CK_RV proxy_C_@\textcolor{red}{\textbf{DigestKey}}@(...){
    /*
    Variable Initialization.
    */
    v5 = (Proxy *)self[1].C_GetSlotInfo;
    v7 = handle;
    result = map_session_to_real(v5,&v7,&map,V3);
    if (!result)
        result = map.funcs->C_@\textcolor{red}{\textbf{DigestKey}}@(v7, mechanism);
    return result;
}
\end{lstlisting}
    \end{minipage}
    \caption{Two Proxy Functions with only distinctions highlighted in red}
    \label{fig:proxy_func}
\end{figure}
\paragraph{\textbf{False Positive Analysis.}}
The false positive outcomes of \toolname{} can be attributed to two primary causes:

\textit{Cause 1: Similar Syntactic Structures of Proxy Functions} - Proxy functions exhibit similar syntactic structures, which can lead to similar semantics. This can make it challenging for Asteria-Pro to differentiate between proxy functions since their semantics are alike. Figure~\ref{fig:proxy_func} provides an illustration of two proxy functions that differ only on line 9. Due to their similar semantics, it becomes difficult to confirm the actual callees, especially when symbols are lacking or when indirect jump tables are involved.

\textit{Cause 2: Compiler-Specific Intrinsic Functions} - Compilers for different architectures utilize various intrinsic functions, which substitute libc function calls with optimized assembly instructions. For example, the gcc-X86 compiler may replace the memcpy function with several memory operation instructions that are specific to the architecture. As a result, the memcpy function may be absent from the list of callee functions used by Asteria's filtering and re-ranking modules. This lack of complete callee function information can lead to a loss of precision in the scoring calculation.

Both causes contribute to the false positive outcomes in \toolname{}, highlighting the challenges in accurately detecting function similarity across different architectures and handling variations in compilers' optimization techniques. Addressing these causes and improving the precision of function similarity detection in such scenarios is an ongoing area of research and development in the field of BCSD.

\begin{table}[h]
\caption{Cross-compiler Evaluation Results.}
\centering
\begin{tabular}{c|lllll}
\hline
\textbf{Metrics} & \textbf{Methods}     & \textbf{gcc-clang} & \textbf{gcc-icc} & \textbf{clang-icc} & \textbf{Avg.} \\
\hline
\multirow{6}{*}{MRR}                 & \textbf{Asteria-Pro} & \textbf{0.755}     & \textbf{0.560}   & \textbf{0.564}     &  \textbf{0.626}             \\
                                     & Asteria              & 0.624              & 0.319            & 0.328              &  0.424             \\
                                     & Trex                 & 0.148              & 0.063            & 0.093               & 0.101              \\
                                     & Gemini               & 0.234              & 0.121            & 0.080              &  0.145             \\
                                     & Safe                 & 0.058              & 0.187            & 0.076              &  0.107             \\
                                     & Diaphora             & 0.727              & 0.370            & 0.384              &  0.494             \\
\hline
\multirow{6}{*}{Recall@Top-1}        & \textbf{Asteria-Pro} & \textbf{0.694}     & \textbf{0.479}   & \textbf{0.486}     & \textbf{0.553}              \\
                                     & Asteria              & 0.541              & 0.244            & 0.256              & 0.347              \\
                                     & Trex                 & 0.099              & 0.031            & 0.040              & 0.057              \\
                                     & Gemini               & 0.164              & 0.079            & 0.048              & 0.097              \\
                                     & Safe                 & 0.027              & 0.152            & 0.031              & 0.070              \\
                                     & Diaphora             & 0.662              & 0.312            & 0.330              & 0.435             \\
\hline
\multirow{6}{*}{Recall@Top-10}       & \textbf{Asteria-Pro} & \textbf{0.864}     & \textbf{0.706}   & \textbf{0.711}     & \textbf{0.760}              \\
                                     & Asteria              & 0.783              & 0.466            & 0.469              & 0.573             \\
                                     & Trex                 & 0.257              & 0.124            & 0.075              & 0.152             \\
                                     & Gemini               & 0.368              & 0.196            & 0.137              & 0.234              \\
                                     & Safe                 & 0.101              & 0.239            & 0.149              & 0.163              \\
                                     & Diaphora             & 0.844              & 0.476            & 0.497              & 0.606             \\
\hline
\end{tabular}
\label{tab:cross_compiler_eva}
\end{table}

\subsubsection{Cross-Comiler Evaluation}

In the cross-compiler evaluation, we conducted experiments using three different compilers: gcc, icc (Version 2021.1 Build 20201112\_000000), and clang (10.0.0), all for the x86 architecture. The evaluation results are presented in Table~\ref{tab:cross_compiler_eva}.
We evaluated the performance of different methods using metrics such as MRR and Recall in the three cross-compiler settings: gcc-clang, gcc-icc, and clang-icc. The average values for all three settings are also provided in the last column of the table.
Our new tool, \toolname{}, consistently outperforms the baseline methods by significant margins across all three compiler combinations. Compared to Asteria, \toolname{} achieves an average improvement of 47.6\% in MRR and Recall, demonstrating its superior performance. The improvements compared to other baseline tools such as Trex, Gemini, Safe, and Diaphora are even more substantial, with average improvements of 596.6\%, 331.7\%, 485.0\%, and 26.7\%, respectively.
It is worth noting that Diaphora achieves surprisingly high precision in the gcc-clang setting, particularly compared to the cross-architecture setting. This may be attributed to the fact that compilers gcc and clang employ similar compilation optimization algorithms, resulting in similar assembly code and abstract syntax tree (AST) structures. However, since Asteria is not trained on a cross-compiler dataset, it exhibits relatively lower precision compared to Diaphora.
Although the precision performances of the methods vary in different compiler combination settings, a consistent trend can be observed. Specifically, higher precision is observed in the gcc-clang setting, while lower precision is observed in the gcc-icc and clang-icc settings, except for Safe. This can be attributed to the fact that the icc compiler employs more aggressive code optimizations, resulting in dissimilar assembly code compared to the other compilers.
Overall, the results of the cross-compiler evaluation demonstrate the effectiveness of \toolname{} in detecting function similarity across different compilers and highlight its superior performance compared to the baseline methods.

\begin{tcolorbox}
    \textbf{Answer to RQ1}: \toolname{} demonstrates superior accuracy in both Task-C and Task-V.
    In Task-C, dominant model in \toolname{} demonstrates the best classification performance by producing the highest AUC (0.997).
    Regarding Task-V, \toolname{} outperforms other baseline methods by a large margin in  MRR, Recall@Top-1, and Recall@Top-10.
    In particular, \toolname{} has 172\%, 236\%, 147\% higher MRR, Recall@Top-1, Recall@Top-10 than the best baseline methods.
    Compared with \methodname{Asteria}, \toolname{} manages to improve it for Task-V with 23.71\% higher MRR, 36.4\% higher Recall@Top-1, and 7.2\% higher Recall@Top-10.
    In a cross-compiler setting, \toolname{} continues to outperform baseline methods by a significant margin. All BCSD methods exhibit higher accuracy in the gcc-clang pairing compared to the gcc-icc or clang-icc pairings, likely because the icc compiler tends to emit highly optimized assembly code.
\end{tcolorbox}

\begin{figure}[h]
    \centering
    \begin{minipage}[t]{0.4\linewidth}
    \includegraphics[width=\linewidth]{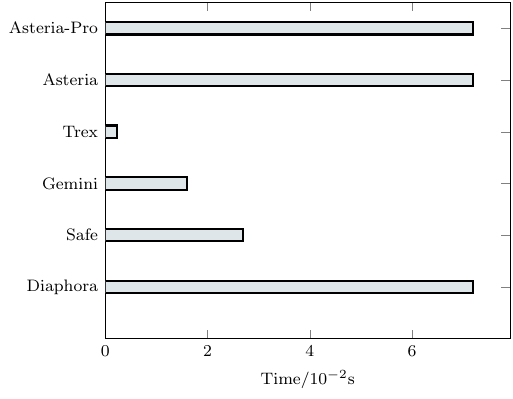}
    \subcaption{Average Feature Extraction Time for One Function}\label{fig:avg_fea_extrac}
\end{minipage}
\begin{minipage}[t]{0.44\linewidth}
        \includegraphics[width=\linewidth]{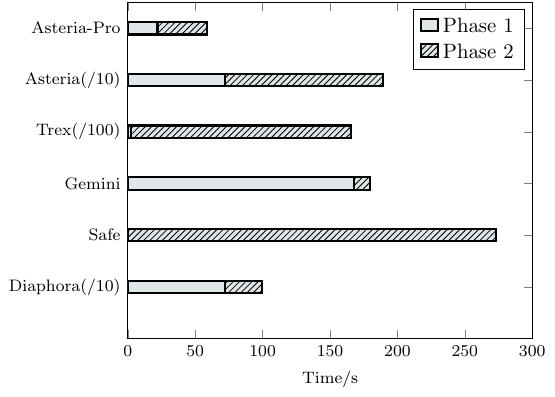}
        \subcaption{Average Time for One Search}\label{fig:avg_time_one_search}
    \end{minipage}
    \caption{Performance Comparison of All Methods on Task-V}
    \label{fig:performance}
\end{figure}

\subsection{Performance Comparison (RQ2)}
In this section, the detection time of function similarity for all baseline approaches and \toolname{} are measured. Since the DK-based prefiltration and DK-based re-ranking modules are intended to enhance performance in Task-V, we only count the timings in Task-V. In task-V, given a source function, methods extract the function features of source and all candidate functions, which is referred to as \textit{phase 1}. Next, the extracted function features are subjected to feature encoding and encoding similarity computation to determine the final similarities, which is referred to as \textit{phase 2}.

As shown in Figure~\ref{fig:avg_fea_extrac}, we calculate the average feature extraction time for each function. The x-axis depicts extraction time, while the y-axis lists various extraction methods.
During feature extraction for one single function, \toolname{}, Asteria, and Diaphora all execute the same operation (i.e., AST extraction), resulting in the same average extraction time.
Since AST extraction requires binary disassembly and decompilation, it requires the most time compared to other methods. 
Trex requires the least amount of time for feature extraction, which is less than 0.001s per function, as code disassembly is the only time-consuming activity.


Figure~\ref{fig:avg_time_one_search} illustrates the average duration of a single search procedure for various methods. The phases 1 and 2 of a single search procedure are denoted by distinct signs. Due to its efficient filtering mechanism, \toolname{} requires the least amount of time (58.593s) to complete a search. Due to its extensive pre-training model encoding computation, Trex is the most time-consuming algorithm. \toolname{} cuts search time by 96.90\%, or 1831.36 seconds, compared to Asteria (1889.96 seconds).

\begin{tcolorbox}
    \textbf{Answer to RQ2:} \toolname{} costs the least average time to accomplish task-V. Compared with Asteria, \toolname{} cuts search time by 96.90\% by introducing the filtering module.
\end{tcolorbox}

\begin{table}[h]
\caption{Accuracy of Different Module Combination}~\label{tab:ablation}
\begin{tabular}{lllll}
\hline
  \textbf{Module Combination} & \textbf{MRR} & \textbf{Recall@Top-1} & \textbf{Recall@Top-10} & \textbf{Average Time(s)} \\ \hline
Pre-filtering + Asteria     & 0.824         & 0.764                 & 0.929                   & 57.8                \\
Asteria + Reranking         & 0.882         & 0.864                 & 0.910                   & 1889.8      \\ \hline
\end{tabular}
\end{table}
\subsection{Ablation Experiments (RQ3)}
To demonstrate the progresses made by different modules of DK-based filtration and DK-based re-ranking, we conduct ablation experiments by evaluating the different module combinations in \toolname{}.
The module combinations are \textit{Pre-filtering + Asteria} and \textit{Asteria + Re-ranking}.
The two module combinations performs Task-V and the results are shown in Table~\ref{tab:ablation}.
For \textit{Asteria + Re-ranking}, the top 20 similarity detection results are re-ranked by the Re-ranking module. 

\subsubsection{Filtration Improvement}
Compared to Asteria, the integration of pre-filtering improves MRR, Recall@Top-1, and Recall@Top-10 by 12.26\%, 16.29\%, and 5.93\%, respectively.
In term of efficiency, it cuts search time by 96.94\%.
The \textit{Pre-filtering + Asteria} combination performs better than \textit{Asteria + Re-ranking} in terms of \textit{Recall@Top-10} and time consumption.
It generates a greater \textit{Recall@Top-10} because it filters out a large proportion of highly rated non-homologous functions. 

\subsubsection{Re-ranking Improvement}
Compared to Asteria, the integration of Re-ranking module improves MRR, Recall@Top-1, and Recall@Top-10 by 20.16\%, 31.51\%, and 3.76\%, respectively.
In terms of efficiency, it costs average additional 0.13s for re-ranking, which is negligible.
Compared to \textit{Pre-filtering + Asteria}, re-ranking module contributes to an increase in \textit{MRR} and \textit{Recall@Top-1} by enhancing the rank of homologous functions.

\begin{table}[t]
    \centering
     \caption{Integration of Pre-Filtration and Re-Ranking Modules with Alternative BCSD Techniques. We denote the integration with BCSD method X as X-I.}
    \label{tab:bcsd_integration}
    \resizebox{\linewidth}{!}{
    \begin{tabular}{c|ll|ll|ll|ll|ll}
    \hline
        Methods & Trex & \textbf{Trex-I} & Gemini & \textbf{Gemini-I} & Safe & \textbf{Safe-I} & Diaphora & \textbf{Diaphora-I}  & Asteria & \textbf{\toolname{}}\\
        \hline
        MRR & 0.310 & \textbf{0.547} & 0.334 & \textbf{0.775} & 0.015 & \textbf{0.533} & 0.020 & \textbf{0.772} & 0.734 & \textbf{0.908} \\
        Recall@Top-1 & 0.192 & \textbf{0.377} &  0.266 & \textbf{0.722} & 0.003 & \textbf{0.484} & 0.018 & \textbf{0.711} & 0.657 & \textbf{0.896} \\
        Recall@Top-10 & 0.579 & \textbf{0.881} & 0.488 & \textbf{0.865} & 0.015 & \textbf{0.603} & 0.027 & \textbf{0.878} & 0.877 & \textbf{0.940}\\
        \hline
    \end{tabular}
    }
\end{table}
\subsubsection{Embedding Baseline Methods.}
We demonstrate the generalizability of innovative BCSD enhancement framework, by integrating our two new components, pre-filtering and re-ranking, with other baseline BCSD methods.
Specifically, we apply all the baseline methods to compute the similarity scores between the remaining functions and the source function after pre-filtering.
Subsequently, we rank the remaining functions in descending order based on their similarity scores and select the top 50 functions for re-ranking.
The final similarity score is obtained by combining the re-ranking score and the score generated by the baseline methods.
Using the final similarity score, we determine the rankings of the top 50 candidate functions and calculate three metrics, namely MRR, Recall@Top-1, and Recall@Top-10.
Table~\ref{tab:bcsd_integration} presents a comparison of the original and integrated versions of the baseline methods, with the baseline method names listed in the first row and their corresponding integrated versions in the next column, such as \textbf{Trex-I} for the integrated version of Trex.
The second to fourth rows provide the values of different metrics, namely MRR, Recall@Top-1, and Recall@Top-10.

The accuracy of the baseline methods is significantly improved by the addition of our two components, with \textbf{Diaphora-I} in particular showing a substantial increase in MRR from 0.02 to 0.772. We manually analyzed the outputs of Diaphora and \textbf{Diaphora-I} to understand the reason for the improved ranking of homologous functions. We found that while Diaphora tends to assign high similarity scores to homologous functions, it also assigns high scores to numerous non-homologous functions, which lowers the ranking of homologous functions.
Specifically, we found that the average score difference between the highest score (i.e., score of top 1) and the score assigned to the homologous function is only 0.11.
By incorporating reranking scores into the final scores, \textbf{Diaphora-I} places a higher emphasis on homologous functions, resulting in improved ranking. 
If homologous functions are present in the top 50 before re-ranking, they are mostly ranked at the top.
\textbf{Safe-I} also shows improved accuracy, although not as substantial as \textbf{Diaphora-I}, as Safe tends to rank homologous functions outside the top 50, reducing the impact of reranking. 

The enhancement framework also effectively enhances the accuracy of Asteria. Specifically, \toolname{} achieves very high MRR and Recall@Top-1, with a notable margin compared to other integrated versions of baseline methods. The high accuracy of \toolname{} enables it to generate more reliable search results, which can significantly reduce the efforts required for vulnerability confirmation when applied to bug search tasks. 
    

\begin{tcolorbox}
    \textbf{Answer to RQ3:} The filtering significantly cuts the calculation time by 96.94\%, and increase precision slightly.
    Re-ranking improves MRR, Recall@Top-1, and Recall@Top-10 by 20.16\%, 31.51\%, and 3.76\%, respectively, with negligible time costs.
    Our enhancement framework, which embeds the BCSD method with pre-filtering and re-ranking modules, has demonstrated a significant improvement in the accuracy of other baseline methods as well.
\end{tcolorbox}

\begin{table}[h]
    \centering
        \caption{Capacity to Filter of Various Filtering Thresholds.}
    \label{tab:filter_thresholds}
    \begin{tabular}{c|cc}
    \hline
     \textbf{$T_{NCL}$}    & \textbf{\# Filtered Function} & \textbf{Recall}  \\
     \hline
     0.1 & 9666.7 & 0.9813\\
     0.2 & 9734.1 & 0.9808\\
     0.3 & 9777.4 & 0.9791\\
     0.4 & 9793.5 & 0.9773 \\
     0.5 & 9805.5 & 0.9737\\
     \hline
    \end{tabular}
\end{table} 

\subsection{Configurable Parameter Sensitivity Analysis (RQ4)}
\toolname{} has two sets of configurable parameters: the filtering threshold $T_{NCL}$ in the pre-filtering algorithm, and the weight values in Equation~\ref{eq:rerank} for the final precision score. In our evaluation, we analyze the impact of these parameters on \toolname{}'s performance by testing different values of $T_{NCL}$ for pre-filtering and varying weight combinations for the final precision score.

\subsubsection{Different Filtering Threshold}\label{sec:threshold_eval}
In Algorithm~\ref{alg:uprelaiton}, the threshold $T_{NCL}$ determines the number of functions that are filtered out. We evaluate the efficacy of the filtering module by utilizing various $T_{NCL}$ values, and the results are presented in Table~\ref{tab:filter_thresholds}. The threshold values range from 0.1 to 0.5 in the first column, where a higher threshold value suggests a more severe selection of the similarity function. The second column indicates the number of functions omitted by the filter, while the third column displays the recall rate in the filteration results. As the threshold value increases, the recall rate declines and the number of filtered-out functions grows. We use 0.1 as our threshold value for two key reasons: a) the high recall rate of filtering results is advantageous for subsequent homologous function detection, and b) there is no significant difference in the number of functions that are filtered out.

\begin{table}[h]
\centering
\caption{Accuracy of \toolname{} with Various Weight Combinations in Equation \ref{eq:rerank}.} \label{tab:weights_sensitivity}
\begin{tabular}{lllll}
\hline
\textbf{$\alpha$}   & \textbf{$\beta$}   & \textbf{MRR}   & \textbf{Recall@Top-1} & \textbf{Recall@Top-10} \\
\hline
0.0          & 1.0          & 0.901          & 0.890                 & 0.930                  \\
\textbf{0.1} & \textbf{0.9} & \textbf{0.908} & \textbf{0.896}        & \textbf{0.940}         \\
0.2          & 0.8          & 0.905          & 0.893                 & 0.938                  \\
0.3          & 0.7          & 0.902          & 0.890                 & 0.937                  \\
0.4          & 0.6          & 0.900          & 0.889                 & 0.935                  \\
0.5          & 0.5          & 0.899          & 0.887                 & 0.934                  \\
1.0          & 0.0          & 0.824          & 0.764                 & 0.929     \\
\hline
\end{tabular}
\end{table}
\subsubsection{Weights in Re-ranking} \label{sec:weights_eva}
 We conducted a sensitivity analysis of different weight values in Equation 16. The evaluation results are presented in Table~\ref{tab:weights_sensitivity}. The first two columns display the combinations of two distinct weights, $\alpha$ and $\beta$, from Equation 16. The last three columns give the values of various metrics, including MRR, Recall@Top-1, and Recall@Top-10.

    We did not test all weight combinations as the accuracy metrics consistently decreased with an increase in $\alpha$. As shown in the table, when $\alpha = 0.1$ and $\beta = 0.9$, Asteria-Pro has the best accuracy. The last column sets $\beta$ to 0.0, meaning the re-ranking score is not included in the final similarity calculation. Therefore, the results are consistent with the combination "Pre-filtering + Asteria" in Section 8.7.

\begin{tcolorbox}
    \textbf{Answer to RQ4:}  The pre-filtering module performs best with a filtering threshold value of 0.1. This threshold allows \toolname{} to filter out 96.67\% of non-homologous functions per search, while achieving a recall rate of 98.13\%.
Regarding the re-ranking score, we found that non-zero weight values have a relatively small effect on the final accuracy of \toolname{}. However, incorporating the re-ranking score significantly improves the precision of the tool.
\end{tcolorbox}

\subsection{Real World Bug Search (RQ5)}
To assess the efficacy of \toolname{}, we conduct a massive real-world search for bugs. To accomplish this, we obtain firmware and compile vulnerability functions to create a firmware dataset and a vulnerability dataset. Utilizing vulnerability dataset, we then apply \toolname{} to detect vulnerable functions in the firmware dataset. To confirm vulnerability in the resulting functions, we design a semi-automatic method for identifying vulnerable functions. Through a comprehensive analysis of the results, we discover intriguing facts regarding vulnerabilities existed in IoT firmware.

\begin{table}[t]
\caption{Vulnerability Dataset}\label{tab:vul_dataset}
\begin{tabular}{l|lll}
\hline
 \textbf{Software} & \textbf{CVE \#} & \textbf{Disclosure Years}      & \textbf{Vulnerable Version Range}                                                                         \\
\hline
\textbf{OpenSSL}  & 22              & 2013$\sim$2016             & \begin{tabular}[c]{@{}l@{}}{[}1.0.0, 1.0.0s{]}\\ {[}1.0.1, 1.0.1t{]}\\ {[}1.0.2, 1.0.2h{]}\end{tabular} \\
\textbf{Busybox}  & 10              & 2015$\sim$2019             & {[}0.38, 1.29.3{]}                                                                                      \\
\textbf{Dnsmasq}  & 14              & 20\{15,17,20,21\}          & {[}2.42, 2.82{]}, 2.86                                                                                  \\
\textbf{Lighttpd} & 10              & 20\{08,10,11,13,14,15,18\} & {[}1.3.11, 1.4.49{]}                                                                                    \\
\textbf{Tcpdump}  & 36              & 2017                       & {[}3.5.1, 4.9.1{]} \\
\hline
\end{tabular}
\end{table}
\subsubsection{Dataset Construction.}
In contrast to our prior work, we expand both the vulnerability dataset and the firmware dataset for a comprehensive vulnerability detection evaluation.
\paragraph{\textbf{Vulnerability Dataset}}
The prior vulnerability dataset of 7 CVE functions is enlarged to \textbf{90}, as shown in Table~\ref{tab:vul_dataset}. Vulnerability information is primarily gathered from the NVD website~\cite{nvd_url}. As shown in the first column, the vulnerabilities are collected from widely used open-source software in IoT firmware, including OpenSSL, Busybox, Dnsmasq, Lighttpd, and Tcpdump. In the second column, the number of software vulnerabilities is listed. In the third column, the timeframe or specific years of the disclosure of the vulnerability are listed. The final column describes the software version ranges affected by vulnerabilities. Note that the version ranges are obtained by calculating the union of all versions mentioned in the vulnerability reports. As a result, \toolname{} is expected to generate vulnerability detection results for all software versions falling within the specified ranges.

\begin{table}[h]
\caption{Firmware Dataset and Its Software Statistics. \textbf{\#} denotes number.}\label{tab:firmware_dataset}
\resizebox{\textwidth}{!}{
\begin{tabular}{c|lll|lllll}
\hline
 & \multicolumn{3}{c}{\textbf{Firmware Dataset}} & \multicolumn{5}{|c}{\textbf{Software Statistics}} \\
 \multirow{-2}{*}{\textbf{Vendor}} & \textbf{Firmware \#} & \textbf{Binary \#} & \textbf{Function \#} & \textbf{OpenSSL} & \textbf{Busybox}& \textbf{Dnsmasq}& \textbf{Lighttpd}& \textbf{Tcpdump}\\
\hline
Netgear         & 548                   & 984                 & 2,627,143               & 349              & 512              & 85               & 14                & 24                                         \\
TP-Link         & 95                    & 177                  & 427,795                 & 66               & 90               & 11               & 3                 & 7                                        \\
Hikvision      & 90                    & 92                   & 279,299                & 55               & 35               & 0                & 0                 & 2                                     \\
Cisco           & 29                    & 66                   & 60,396                 & 23               & 26               & 10               & 5                 & 2                                       \\
Schneider       & 10                    & 20                   & 31,228                & 7                & 9                & 2                & 2                 & 0                                       \\
Dajiang         & 7                     & 16                   & 57,275                 & 7                & 7                & 1                & 0                 & 1                 \\          \hline  
All             & 779                   & 1,355                  & 3,483,136            &507                & 679               & 109           & 24                & 36               \\
\hline
\end{tabular}
}
\end{table}

\paragraph{\textbf{Firmware Dataset}}
We download as much of firmware from six popular IoT vendors as we could, consisting of Netgear~\cite{url:netgear}, Tp-Link~\cite{tplink-url}, Hikvision~\cite{hikvision-url}, Cisco~\cite{cisco-url}, Schneider~\cite{schneider-url}, and Dajiang~\cite{dajiang_url} as shown in first column of Table~\ref{tab:firmware_dataset}.
These firmware are utilized by routers, IP cameras, switches, and drones, all of which play essential parts in our life.
The second column shows the firmware numbers, which range from 7 to 548.
The third and fourth columns gives numbers of binaries and functions after unpacking firmware by using binwalk.
Note that the binary number is the number of software selected to be in the vulnerability dataset.
The fifth column to ninth column gives the five software numbers in all firmware vendors.
OpenSSL and Busybox are widely integrated in these IoT firmware as their numbers are close to those of the firmware.
Through querying their official websites for device type information, we find that the majority of Hikvision vendor firmware is for IP cameras, whereas Cisco vendor firmware is for routers. 
In particular, IP camera firmware incorporates less software than router firmware because routers offer more functionality.
For example, the firmware of the Cisco \textit{RV340} router includes OpenSSL, Tcpdump, Busybox, and Dnsmasq, whereas the majority of IP camera firmware only include OpenSSL.
Similarly, the majority of the firmware of Netgear and Tp-Link consists of routers, while Schneider and Dajiang'firmware include specialized devices such as Ethernet Radio and Stabilizers.

\subsubsection{Large Scale Bug Search.}
\toolname{} is employed to identify vulnerable homologous functions among 3,483,136 firmware functions by referencing 90 functions from the vulnerability dataset.
Specifically, in order to expedite the detection process, vulnerability detection is restricted to the same software between firmware dataset and vulnerability dataset. 
For instance, the vulnerable functions disclosed in OpenSSL are utilized to detect vulnerable homologous functions in OpenSSL in the firmware dataset.
For each software $S$, we first extract features (i.e., ASTs and call graphs) of all functions in firmware dataset and vulnerable functions in vulnerability dataset.
For each vulnerability disclosed in $S$, the pre-filtration module uses the call graph to filter out non-homologous functions, followed by the Tree-LSTM model encoding all remaining functions as vectors. 
\toolname{} then computes the AST similarity between the vulnerable function vectors and the firmawre function vector. 
\toolname{} computes reranking scores based on the top 20 of AST similarities based on similarity scores, since the evaluation demonstrates a very high recall in the top 20. 
As a final step, \toolname{} generates 20 candidate homologous functions for each $S$ as a \textbf{bug search result} for each vulnerability.
To further refine the bug search results, we compute the average similarity score of homologous functions in Section~\ref{sec:rq1_eval} and use it to eliminate non-vulnerable functions.
In particular, the average similarity score of $0.89$ is used to eliminate 3987 of 5604 results.
We perform heuristic confirmation of vulnerability for the remaining 1617 results.

\paragraph{\textbf{Vulnerability Confirmation Method}}
We devise a semi-automatic method for confirming the actual vulnerable functions from the candidate homologous functions. 
The method makes use of the symbols and string literals within the firmware binaries of the target. 
Specifically, we use unique regular expressions to match version strings for each software and to extract function symbols from the software.
The method is then comprised of two distinct operations that correspond to two distinct vulnerable circumstances $VC_1$, $VC_2$.
\begin{itemize}
    \item \textbf{$VC_1$.} In this circumstances, the target binary contains version string (e.g., ``OpenSSL 1.0.0a'') and the symbol of target function is not removed.
    \item \textbf{$VC_2$.} Target binary contains version strings whereas the symbol of vulnerable homologous function is removed.
\end{itemize}
The versions of software listed in Table~\ref{tab:vul_dataset} are easy to extract using version strings~\cite{duan2017identifying}.
The descriptions of the two confirmation operations $CO_1$ and $CO_2$ are as follows:
\begin{itemize}
    \item \textbf{$CO_1$.} For \textbf{$VC_1$}, we confirm the vulnerable function based on the version and name of the target software. 
In particular, a vulnerable function is confirmed when the following two conditions are met: 1) software version is in vulnerable version range, 2) the vulnerable function name retains after elimination with average similarity score.
    \item \textbf{$CO_2$.} For \textbf{$VC_2$}, if the software versions are in the range of vulnerable versions, we manually compare the code between the CVE functions and remaining functions to confirm the vulnerability.
\end{itemize}
\begin{table}[t]
\caption{Numbers of Vulnerable Functions, Software, Firmware in Confirmation Results.}\label{tab:vul_numbers}
\resizebox{\textwidth}{!}{
\begin{tabular}{l|llllll|llllll|l}
\hline
 \multirow{2}{*}{\textbf{vendor}} & \multicolumn{6}{c|}{\textbf{Vulnerable Function \#}}                                                          & \multicolumn{6}{c|}{\textbf{Vulnerable Software \#}}                                                          & \multicolumn{1}{c}{\multirow{2}{*}{\textbf{\begin{tabular}[c]{@{}c@{}}Vulnerable\\  Firmware \#\end{tabular}}}} \\
\cline{2-13}
     & \textbf{OpenSSL} & \textbf{Busybox} & \textbf{Dnsmasq} & \textbf{Lighttpd} & \textbf{Tcpdump} & \textbf{All} & \textbf{OpenSSL} & \textbf{Busybox} & \textbf{Dnsmasq} & \textbf{Lighttpd} & \textbf{Tcpdump} & \textbf{All} & \multicolumn{1}{c}{}                                                                                           \\\hline
Netgear                                & 367              & 0                & 31               & 0                 & 26               & 424          & 133              & 0                & 7                & 0                 & 10               & 150          & 145 (26.46\%)                                                                                                  \\
TP-Link                                & 394              & 9                & 0                & 2                 & 5                & 410          & 36               & 3                & 0                & 2                 & 5                & 46           & 36 (37.89\%)                                                                                                   \\
Hikvision                              & 553              & 0                & 0                & 0                 & 12               & 565          & 52               & 0                & 0                & 0                 & 1                & 53           & 53 (58.89\%)                                                                                                   \\
Cisco                                  & 0                & 0                & 0                & 0                 & 2                & 2            & 0                & 0                & 0                & 0                 & 2                & 2            & 2 (6.90\%)                                                                                                     \\
Schneider                              & 10               & 0                & 0                & 0                 & 0                & 10           & 1                & 0                & 0                & 0                 & 0                & 1            & 1 (10.00\%)                                                                                                    \\
Dajiang                                & 70               & 0                & 0                & 0                 & 1                & 71           & 7                & 0                & 0                & 0                 & 1                & 8            & 7 (100.00\%)    \\
\hline
Total  & 1,394 & 9 & 31 & 2 & 46 & 1,482 & 229 & 3 & 7 & 2 & 19 & 260 & 244 (31.32\%)
\\\hline
\end{tabular}
}
\end{table}

\begin{figure}[t]
    \centering
    \includegraphics[width=\linewidth]{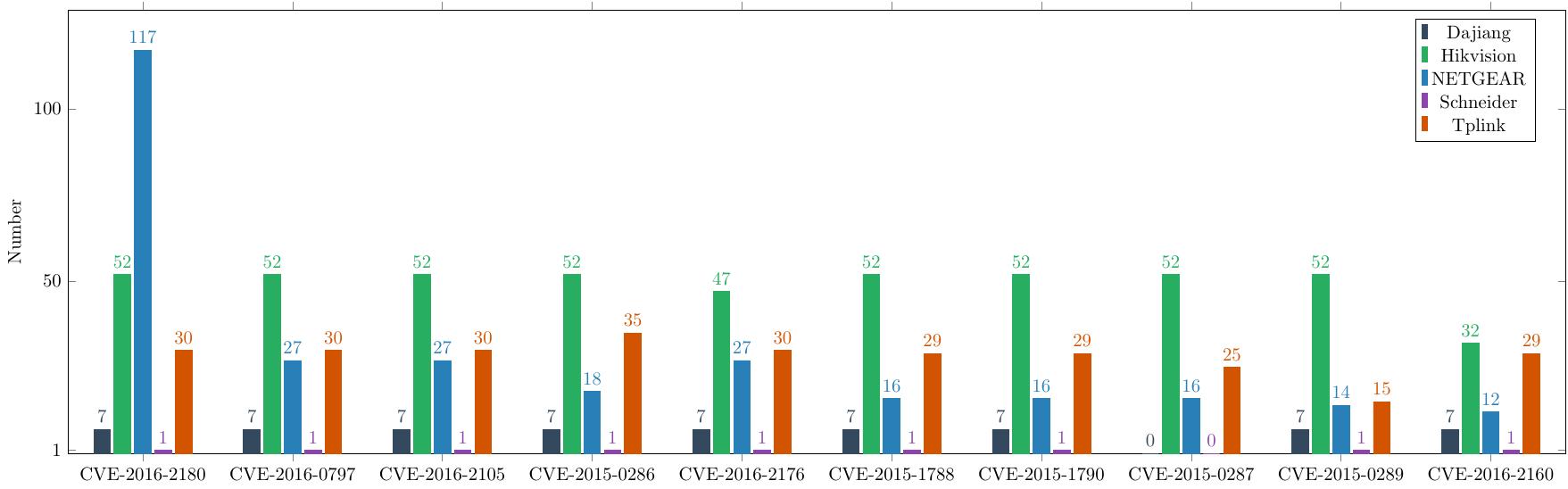}
    \caption{\# of Vulnerable Functions Detected from Five Vendors}
    \label{fig:vul_detected}
\end{figure}
\begin{figure}[h]
    \centering
    \begin{minipage}{0.33\linewidth}
        \includegraphics[width=\linewidth]{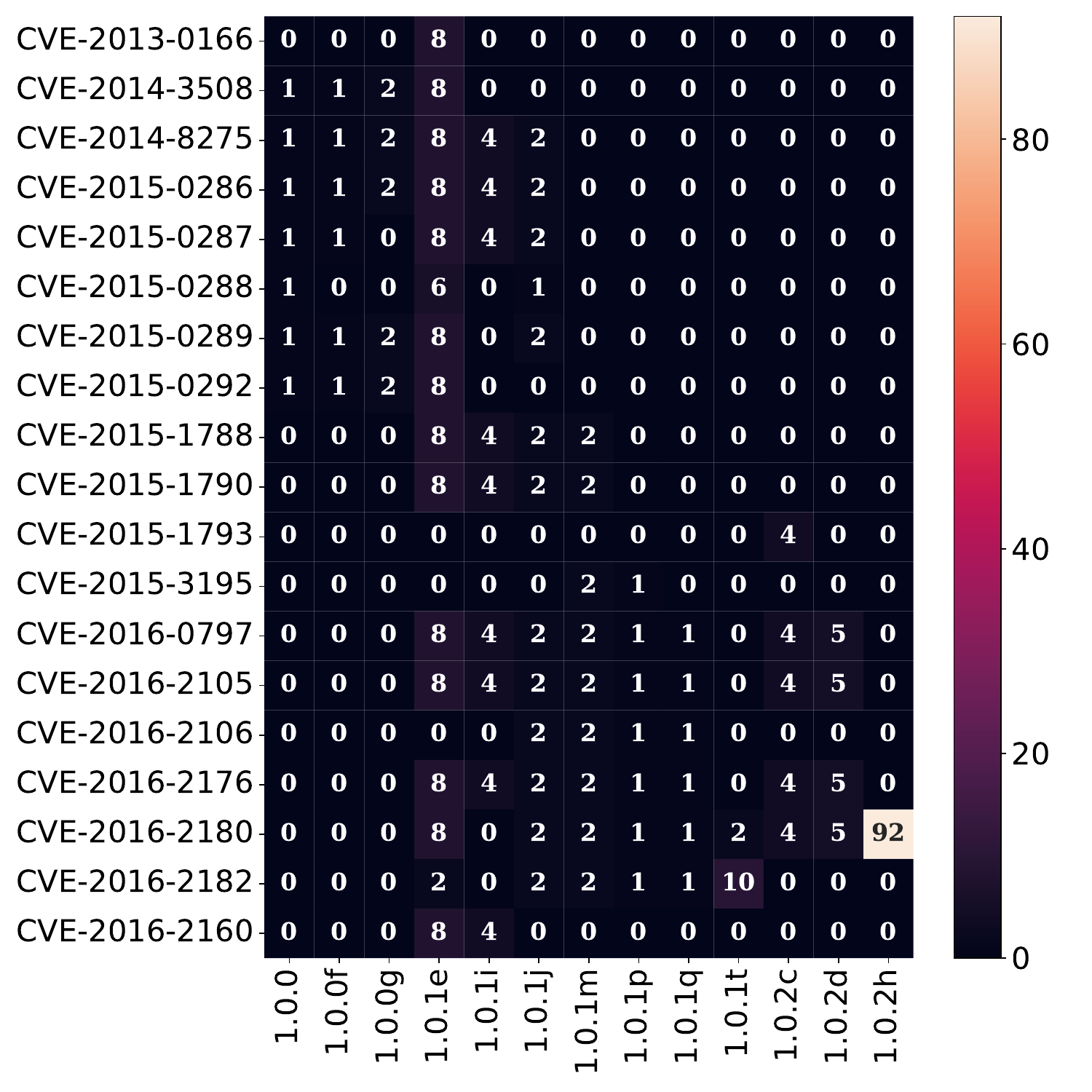}
        \subcaption{Netgear}
    \end{minipage}
    \begin{minipage}{0.33\linewidth}
        \includegraphics[width=\linewidth]{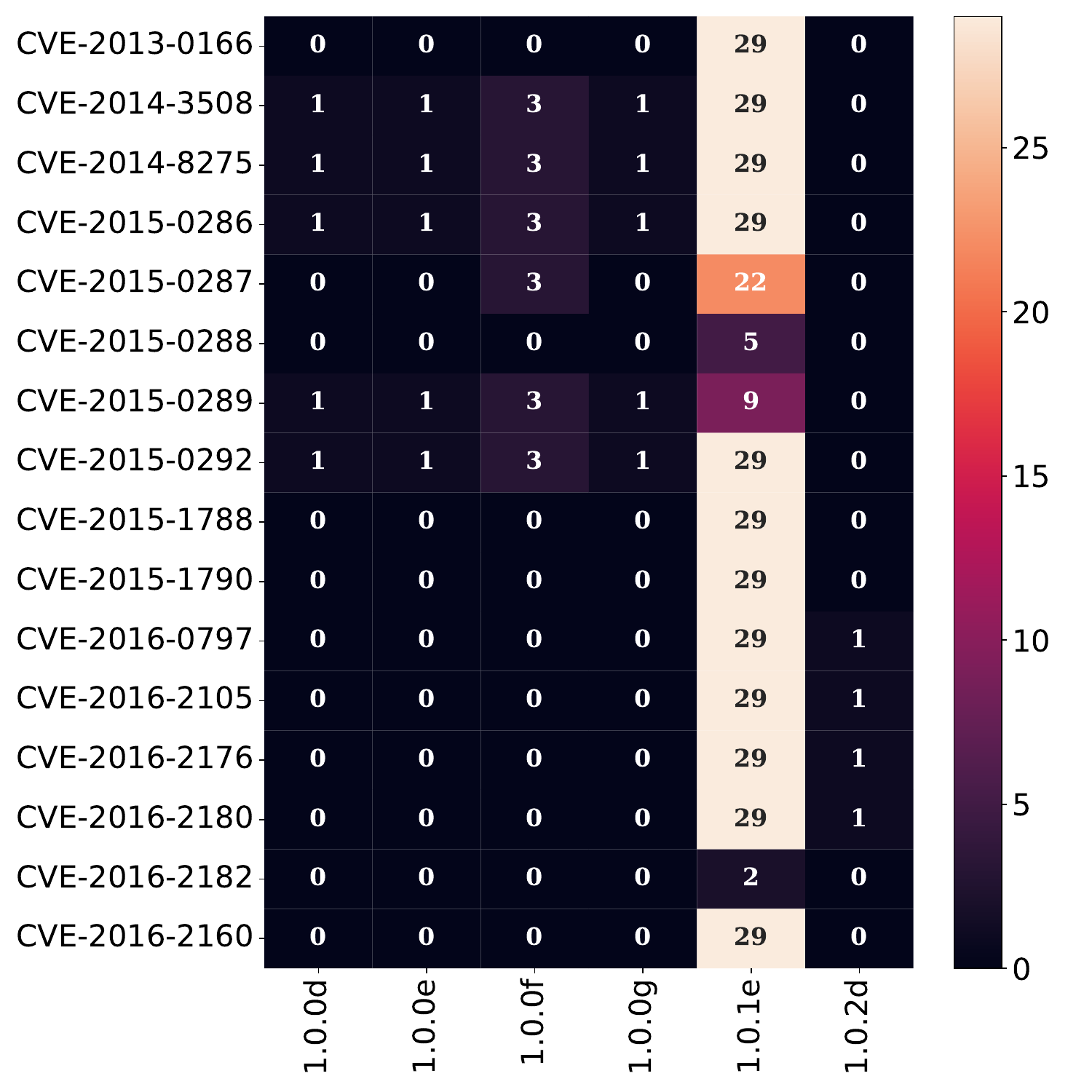}
        \subcaption{TP-Link}
    \end{minipage}
    \begin{minipage}{0.33\linewidth}
        \includegraphics[width=\linewidth]{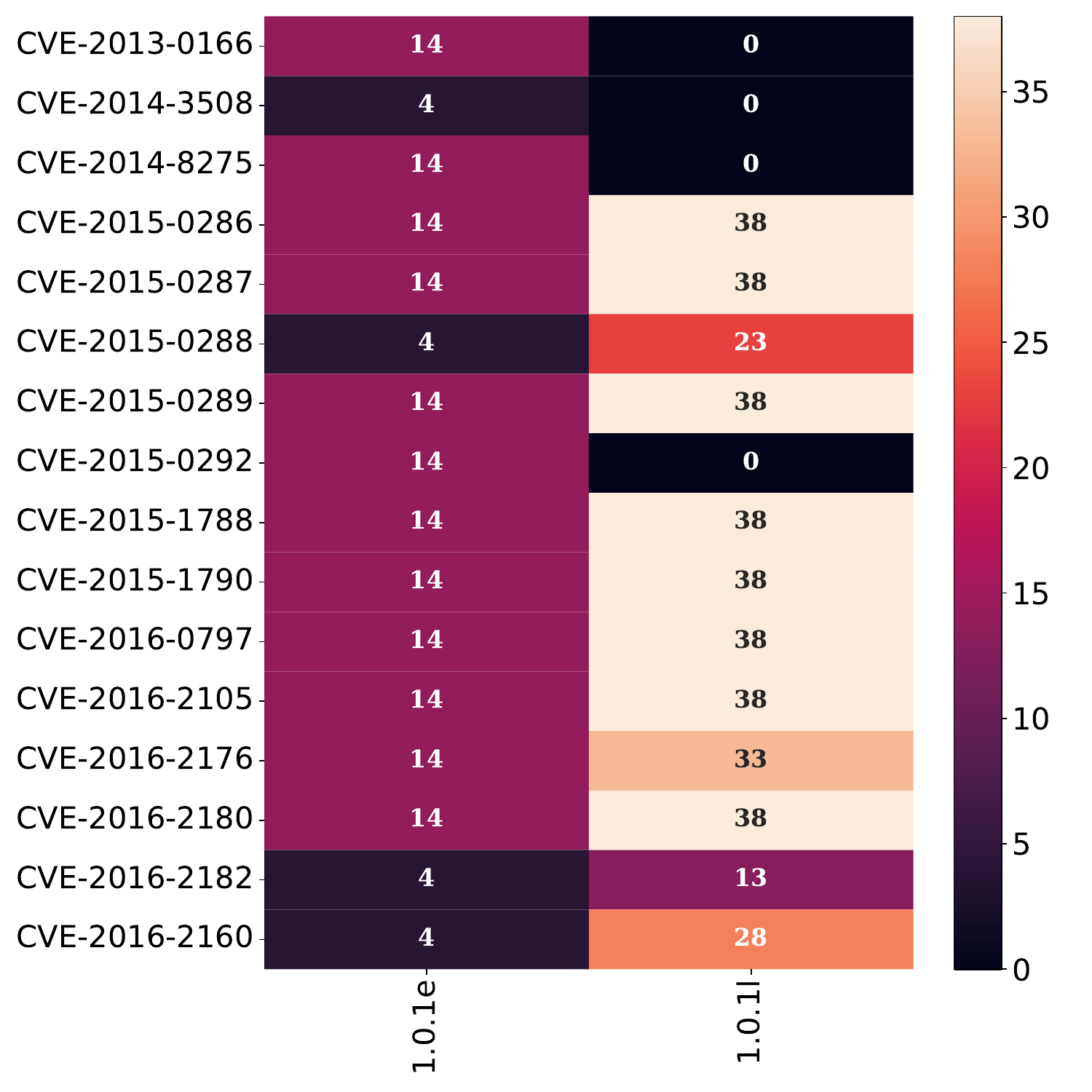}
        \subcaption{Hikvision}
    \end{minipage}
    \caption{The Distribution of CVEs for Different OpenSSL Versions in vendors Netgear, TP-Link, Hikvision from left to right.}
    \label{fig:heat_map_cve}
\end{figure}
\paragraph{\textbf{Results Analysis}}
In Table~\ref{tab:vul_numbers}, we tally the number of vulnerable functions, software, and firmware upon vulnerability confirmation. The first column contains the names of different vendors. The second through sixth columns show the amount of vulnerable functions in various software, while the seventh column indicates the total number of vulnerable functions across all vendors. The eighth through twelfth columns display the amount of vulnerable software binaries in various software, while the thirteenth column provides the total number of vulnerable software binaries. 
According to the seventh column of Table~\ref{tab:firmware_dataset}, there are a total of \textbf{1,482} vulnerable functions. 
1456 are confirmed by $CO_1$, whereas 26 are confirmed by $CO_2$. 
For a total of 1,456 $CO_1$ vulnerable functions, 1,377 vulnerable functions rank first and 79 vulnerable functions rank second. 
$CO_2$ is performed on 47 detection results, of which 26 are confirmed. 
 the 21 unconfirmed detection results can be attributed to two reasons. Firstly, 18 of them were due to the fact that the target binaries detected did not contain any of the target vulnerable functions. For example, we were unable to detect the vulnerable function `EVP\_EncryptUpdate' of CVE-2016-2106 from the `libssl.so' library of OpenSSL since it exists in the `libcrypto.so' library. 
 Secondly, 3 of the unconfirmed results were ranked in the top 20 but were subsequently filtered out by the similarity threshold  used in real world bug search setting. 
A large proportion of vulnerable functions are found in the OpenSSL software used by the three vendors. 
The number of vulnerable software is consistent with this circumstance. 
The final column shows the number of firmware containing at least one vulnerable function, together with its proportion of total firmware. 
Every Dajiang firmware contains at least one CVE vulnerability because all OpenSSL components used in firmware are vulnerable. 
In addition, Hikvision is detected to have a large proportion of vulnerable firmware (58.89\%). 
To inspect the CVE vulnerable function distribution, we plot the top 10 CVEs and their distributions in five vendors except Cisco in Figure~\ref{fig:vul_detected}, since Cisco takes additional two CVEs.
\begin{itemize}
    \item \textbf{Top 10 CVE Analysis.} Figure~\ref{fig:vul_detected} demonstrates the top 10 CVE distribution in various vendors.
    The total number of discovered CVE vulnerabilities decreases from left to right along the x-axis.
    Except for CVE-2015-0287, all of the top 10 CVE vulnerabilities are discovered in every \textit{Dajiang} firmware.  
    This is because Dajiang utilizes an outdated version of OpenSSL 1.0.1h that contains numerous vulnerable functions~\cite{openssh_1_0_1h}. 
    Although Hikvision firmware has the third largest number of firmware, it has the most vulnerable functions in our experiment settings. 
    The reason for this is that Hikvision firmware heavily uses OpenSSL-1.0.1e (184) and OpenSSL-1.0.1l (401) versions, both of which contain a large number of vulnerabilities.
    \textbf{Finding:} Since they typically adopt the same vulnerable software version, it is highly plausible that firmware from the same vendor and released at the same period contains identical vulnerabilities. Security analysts can quickly narrow down the vulnerability analysis based on the firmware release date.

    \item \textbf{CVE and Version Analysis.} Figure~\ref{fig:heat_map_cve} depicts the distribution of vulnerable OpenSSL versions for various CVEs from various vendors. where the x-axis represents the version and the y-axis represents the CVE ID associated with the vulnerability.
Each square in each subfigure indicates the number of OpenSSL versions that are vulnerable and contain the corresponding CVE along the y-axis.
The number is greater the lighter the red colour.
The left subfigure demonstrates that OpenSSL 1.0.2h is widely used by Netgear, resulting in a significant number of CVE-2016-2180 vulnerabilities (92).
Additionally, OpenSSL version 1.0.1e exposes the majority of CVEs listed on the y-axis, which may increase the device's attack surface.
The TP-Link firmware incorporates OpenSSL version 1.0.1e, resulting in brighter hues.
Hikvision firmware utilizes versions 1.0.1e and 1.0.1l, which are vulnerable to a number of CVEs.
Comparing vulnerability distribution in OpenSSL version 1.0.1e among different vendors reveals inconsistencies in the existence of vulnerabilities.
For instance, CVE-2016-2106 is present in OpenSSL 1.0.1e from Hikvision but not from Netgear and TP-Link.
\textbf{Finding:} Despite using the same version of software, various vendor firmware behaves differently in terms of vulnerability since they can tailor the software to the device's specific capabilities.

    \item \textbf{CVE-2016-2180 Analysis.} The CVE-2016-2180, which is a remote \textit{Denial-of-Service} flaw caused by received forged time-stamp file, impacting OpenSSL 1.0.1 through 1.0.2h, exists in 207 firmware. 
    NETGEAR is responsible for 117 of these, as it deploys 92 OpenSSL 1.0.2h out of a total of 548 firmware. 
    NETGEAR incorporated an extra nine OpenSSL 1.0.2 series software and sixteen OpenSSL 1.0.1 series software. 
    The vulnerable version 1.0.2h was released in May 2016, and by comparing their timestamps, we determined that OpenSSL 1.0.2h was integrated into firmware between 2016 and 2019. \textbf{Finding:} Even after their vulnerabilities have been discovered, the vulnerable versions of software continue to be used for firmware development. 
\end{itemize}

Based on the confirmation results, \toolname{} manages to detect 1,482 vulnerable functions out of 1617 bug search results, indicating that \toolname{} achieves a high vulnerability detection precision of \textbf{91.65\%} under our experiment settings.
By randomly selecting 1,000 of 5,604 bug search results, we manually validate the existence of vulnerabilities in software binaries in order to calculate the recall.
Among 1000 bug search results, 205 target function are confirmed to be vulnerable by checking software versions and the vulnerable functions.
Targeting 205 vulnerable functions, \toolname{} detects 53 of them, representing a recall rate of \textbf{$25.85\%$}.

\begin{figure}[h]
    \centering
    \begin{minipage}{0.52\linewidth}
\begin{lstlisting}[language=C, escapechar=@,breaklines=true, breakatwhitespace=false, linewidth=8.4cm]
{
...
if (ndo->ndo_uflag) {
    u_int i;
    char const *sep = "";
    @\colorbox{red}{ND\_PRINT((ndo, " fh["));}@
    for (i=0; i<len; i++) {
        @\colorbox{orange}{ND\_PRINT((ndo, "\%s\%x", sep, dp[i]));}@
        @\colorbox{orange}{sep = ":";}@
    }
    ND_PRINT((ndo, "]"));
    return;
}
@\colorbox{yellow}{Parse\_fh((const u\_char *)dp, len, \&fsid, \&ino,}@
      @\colorbox{yellow}{NULL, \&sfsname, 0);}@
@\colorbox{yellow}{if (sfsname) \{}@
    @\colorbox{yellow}{...}@
    @\colorbox{yellow}{strncpy(temp, sfsname, stringlen);}@
    @\colorbox{yellow}{temp[stringlen] = '\\0';}@
    @\colorbox{yellow}{spacep = strchr(temp, ' ');}@
    @\colorbox{yellow}{if (spacep)}@
        @\colorbox{yellow}{*spacep = '\\0';}@
    @\colorbox{yellow}{ND\_PRINT((ndo, " fh \%s/", temp));}@
} else {
    @\colorbox{green}{ND\_PRINT((ndo, " fh \%d,\%d",}@
      @\colorbox{green}{fsid.Fsid\_dev.Major, fsid.Fsid\_dev.Minor));}@
}
@\colorbox{cyan}{if(fsid.Fsid\_dev.Minor $==$ 257)}@
    @\colorbox{cyan}{ND\_PRINT((ndo, "\%s", fsid.Opaque\_Handle));}@
@\colorbox{cyan}{else}@
    @\colorbox{cyan}{ND\_PRINT((ndo, "\%ld", (long) ino));}@
}
\end{lstlisting}
      \subcaption{Source Code of Vulnerable Function.}
    \end{minipage}
        \begin{minipage}{0.46\linewidth}
\begin{lstlisting}[language=C, escapechar=@,breaklines=true, breakatwhitespace=false,linewidth=7.7cm]
{
...
if ( v8 )
  {
    @\colorbox{red}{printf(" fh[");}@
    if ( v4 ){
      for ( i = 0; i < v4; ++i ){
        v11 = *(_DWORD *)&v3[4 * i];
        @\colorbox{orange}{printf("\%s\%x", v9, v11);}@
        @\colorbox{orange}{v9 = ":";}@
      }
    }
    putchar(93);
    return &v3[v5];
  }else{
    @\colorbox{yellow}{Parse\_fh(v3, v4, v13, \&v16, 0, \&src, 0);}@
    @\colorbox{yellow}{if ( src )\{}@
      @\colorbox{yellow}{strncpy(temp\_5711, src, 0x40u);}@
      @\colorbox{yellow}{byte\_B9E04 = 0;}@
      @\colorbox{yellow}{v12 = strchr(temp\_5711, 32);}@
      @\colorbox{yellow}{if ( v12 )}@
        @\colorbox{yellow}{*v12 $=$ 0;}@
      @\colorbox{yellow}{printf(" fh \%s/", temp\_5711);}@
    }else{
      @\colorbox{green}{printf(" fh \%d,\%d/", v13[1], v13[0]);}@
    }
    @\colorbox{cyan}{if ( v13[0] $==$ 257 )}@
      @\colorbox{cyan}{printf("\%s", v14);}@
    @\colorbox{cyan}{else}@
      @\colorbox{cyan}{printf("\%ld", v16);}@
    return &v3[v5];
  }
\end{lstlisting}
         \subcaption{Part of Decompiled Code in Detected Function.}
    \end{minipage}
    \caption{Inlined Vulnerable Function in Detected Function. The semantics of the code with the same background color are same.}
    \label{fig:inlined_bug_code_example}
\end{figure}
\paragraph{\textbf{Finding Inlined Vulnerable Code.}}
During the analysis of mismatched cases, in which the target homologous functions are not in the top ranking position, we observe that the top-ranked functions contain the same vulnerable code.
We use CVE-2017-13001 as an illustration of inlined vulnerable code detection.
CVE-2017-13001 is a buffer over-read vulnerability in the Tcpdump \textit{nfs\_printfh} function prior to version 4.9.2.
After a confirmation operation $CO_2$, \toolname{} reports a single function, \textit{parsefh} as being vulnerable.
We manually compare the decompiled code of the \textit{parsefh} function to the source code of \textit{nfs\_printfh} in tcpdump version 4.9.1 (i.e., vulnerable version).
Figure~\ref{fig:inlined_bug_code_example} demonstrates that the source code of \textit{nfs\_printfh} (on the left) and the partial code of \textit{parsefh} (on the right) are consistent.
We designate codes with apparently identical semantics with distinct backdrop hues.
In other words, during compilation, function \textit{nfs\_printfh} is inlined into function \textit{parsefh}.
As a result, the function \textit{parsefh} contains CVE-2017-13001 vulnerable code, and \toolname{} manages to identify the inlined vulnerable code.
\toolname{} has detected an additional eight instances of inlined vulnerable code out of 20 functions in vulnerable circumstance $VC_2$.

The preceding analysis and conclusions are constrained by the dataset we constructed, which offers security analysts some recommendations for the security analysis of firmware.


\begin{tcolorbox}
    \textbf{Answer to RQ5.} We employ 90 CVE vulnerabilities to search for bugs in 3,483,136 real firmware functions.
\toolname{} detects 1,482 vulnerable functions with a high level of precision of 91.65\%.
In addition, the capability of \toolname{} to identify inlined vulnerable code is stated and illustrated in detail.
In conclusion, \toolname{} generates bug search results with a high degree of confidence, thereby reducing analysis labor by a substantial margin.
\end{tcolorbox} 

\section{Threats to Validation.}
Threats to internal validity come from these aspects.
\begin{itemize}
    \item We use vulnerable version ranges collected from the NVD website to aid vulnerability confirmation in a real-world bug-search experiment. The vulnerable version ranges may be inaccurate, and vulnerabilities may be missed or incorrectly stated.
    We will conduct additional verification of the susceptible version ranges by confirming the existence of vulnerable code.
    \item We adopt IDA Pro to decompiling and generate ASTs from the functions in binaries. As pointed by study~\cite{liu2020far}, accurate decompiling and binary analysis is not easy. The errors in AST generation may affect the AST similarity calculation and further affect the results.
\end{itemize}

Threats to external validity rise from following issue.

\begin{itemize}
    \item In practice, firmware binaries may be compiled with distinct compiler (e.g., clang), compiler version, and optimization level for special-purpose compilation.
Different compilation configurations alter the AST structures and call graphs, sometimes leading in lower scores for homologous functions.

\end{itemize}

\section{Discussion}

\subsection{How does the re-ranking module solve the function inline issues?}
Inlining functions is a common optimization technique used by compilers to improve the performance of code execution. The decision of whether or not to inline a function is based on various factors, including the size of the function, the frequency of function calls, and the complexity of the code. In general, inlining smaller functions tends to be more beneficial than inlining larger functions.

When smaller functions being inlined, the re-ranking module in Asteria-Pro will be capable of handling inline issues by considering both function similarities and the match of callee relational structures.
Specifically, the module matches all callee functions between the source function and target functions, which allows for high similarity even if one callee function is inlined to the target function. This is because the target function can still maintain a relatively high similarity with the source function even after incorporating inlined code, and the un-inlined callee functions still contribute to the final similarity score. As a result, Asteria-Pro exhibits high metric values (i.e., recall and MRR) in our evaluation based on the contribution of the target function code and all its callee functions. Through manual analysis of search results in real-world bug detection, we have also demonstrated that Asteria-Pro is capable of finding homologous vulnerable functions that contain inlined function code.


\subsection{What is the design difference between pre-filtering, re-ranking and SCA tools.}
Some software composition analysis (SCA) tools have adopted similar feature when comparing to the pre-filtering and re-ranking module of \toolname{}.
For example, Modx~\cite{yang2022modx} matches string literals and whole call graph between two libraries, and LibDB~\cite{tang2022libdb} adopts string literals and exported function names to measure the similarity of libraries.
The usages of string literals and call graphs are quite straightforward.

However, we would like to highlight some conceptual-level differences between our approach and these prior works. While the use of string literals and call graphs is indeed straightforward, it can be challenging to apply them to function matching, particularly when functions lack string literals or are leaf nodes in call graphs. Additionally, callee functions of a target function may not be exported functions, meaning that function names are removed and cannot be used for matching.

To address these challenges, our approach differs from SCA methods in several key ways. Firstly, we utilize the local context extracted from call graphs in both the pre-filtering and re-ranking modules to efficiently remove non-homologous functions and confirm homologous ones. Secondly, we introduce an algorithm called ``UpRelation" to utilize caller relations from call graphs in pre-filtering. The algorithm leverages the \textit{genealogist} of parent nodes to identify potential homologous functions. It achieves this by matching the \textit{genealogist} of parent nodes and retaining the child nodes of the matched parent nodes. This approach is particularly useful when the target function is a leaf node in the call graph and does not contain any string literals. Lastly, our re-ranking module considers both structural and semantic similarities of functions, resulting in more accurate ranking of homologous functions. Specifically, the re-ranking module uses Asteria to calculate similarities when callee functions are not exported functions.

\subsection{What will \toolname{} perform on cross-optimization settings?}
Although we did not evaluate the performance of \toolname{} in cross-optimization settings, it is worth discussing the potential impact of such settings on the performance of our method. 
Cross-optimization refers to the situation where the training and testing sets are compiled with different optimization settings. 
This is a common scenario in practice as different developers may use different optimization flags, or the same developer may use different optimization levels for different releases. 
Previous studies have shown that cross-optimization can significantly affect the accuracy of BCSD methods, as the semantic features extracted from the binary code may change depending on the optimization settings. For instance, in a study by Liu et al.~\cite{liu2018alphadiff}, the accuracy of a state-of-the-art BCSD method dropped from 95.3\% to 46.2\% when tested in cross.

In the case of our method, \textbf{Asteria-Pro}, which is based on the Tree-LSTM architecture, the impact of cross-optimization on its performance is likely to be substantial. 
This is because Tree-LSTM model is sensitive to AST structure and summarizes semantics by identifying structure patterns. 
Therefore, if the source and target functions are compiled with different optimization settings, the Tree-LSTM may not be capable to summarize the expected semantic from substantial AST structure transformation and thus produce inaccurate results.

Moreover, training our model on cross-optimization settings would require significant computational resources and time, which may not be feasible in practice. 
Therefore, we have not evaluated our method on cross-optimization settings in this study. Nevertheless, we acknowledge that cross-optimization is an essential consideration for evaluating \textbf{Asteria-Pro}'s generalizability, and we encourage future studies to investigate this aspect further.

\section{Related Works}

\subsection{Feature-based Methods}
When considering the similarity of binary functions, the most intuitive way  is to utilize the assembly code content to calculate the edit distance for similarity detection between functions.
Khoo \etal{} concatenated consecutive mnemonics from assembly language into the N-grams for similarity calculation~\cite{khoo2013rendezvous}.
David \etal{} proposed Trecelet, which concatenates the instructions from adjacent basic blocks in CFGs for similarity calculation~\cite{David:Tracelet}.
Saebjornsen \etal{} proposed to normalize/abstract the operands in instructions, \eg{} replacing registers such as eax or ebx with string ``reg'', and conduct edit distance calculation based on normalized instructions~\cite{Saebjornsen:2009}.
However, binary code similarity detection methods based on disassembly text can not be applied to cross-architecture detection since the instructions are typically different in different architectures.
The works in \cite{pewny2015cross}, \cite{chandramohan2016bingo}, \cite{eschweiler2016discovre}, \cite{xue2018accurate} utilize cross-architecture statistical features for binary code similarity detection.
Eschweiler \etal{}~\cite{eschweiler2016discovre} defined statistical features of functions such as the number of instructions, size of local variables.
They utilized these features to calculate and filter out candidate functions.
Then they performed a more accurate but time-consuming calculation with the graph isomorphism algorithm based on CFGs.
Although this method takes a pre-filtering mechanism, the graph isomorphism algorithm makes similarity calculation extremely slow.
To improve the computation efficiency, Feng \etal{} proposed {Genius} which utilizes machine learning techniques for function encoding~\cite{feng2016scalable}.
{Genius} uses the statistical features of the CFG proposed in~\cite{eschweiler2016discovre} to compose the attributed CFG (ACFG).
Then it uses a clustering algorithm to calculate the center points of ACFGs and forms a codebook with the center points.
Finally, a new ACFG is encoded into a vector by computing the distance with ACFGs in the codebook and the similarity between ACFGs is calculated based on the encoded vectors.
But the codebook calculation and ACFG encoding in {Genius} are still inefficient.
Xu \etal{} proposed \gemini{} based on Genius to encode ACFG with a graph embedding network~\cite{xu2017neural} for improving the  accuracy and efficiency.
However, the large variance of binary code across different architectures makes it difficult to find architecture-independent features~\cite{d2015correctness}.

\subsection{Semantic-based Methods}
For more accurate detection, semantic-based features are proposed and applied for code similarity detection.
The semantic-based features model the code functionality, and are not influenced by different architectures.
Khoo \etal{} applied symbolic execution technique for detecting function similarity~\cite{Luo:SoftwarePlagiarism}.
Specifically, they obtained input and output pairs by executing basic blocks of a function.
But the input and output pairs can not model the functionality of the whole function accurately.
Ming \etal{} leveraged the deep taint and automatic input generation to find semantic differences in inter-procedural control flows for function similarity detection~\cite{ming2012ibinhunt}.
Feng~\etal{} proposed to extract conditional formulas as higher-level semantic features from the raw binary code to conduct the binary code similarity detection~\cite{feng2017extracting}.
In their work, the binary code is lifted into a platform-independent intermediate representation (IR), and the data-flow analysis is conducted to construct formulas from IR.
Egele \etal{} proposed the blanket execution, a novel dynamic equivalence testing primitive that achieves complete coverage by overwriting the intended program logic to perform the function similarity detection~\cite{BlanketExecution}.
These semantic-based features capture semantic functionalities of a function to reduce the false positives.
Pei \etal{} proposes Trex~\cite{trex}, applying a transfer-learning-based framework to automate learning execution semantics from functions' micro-traces, which are forms of under-constrained dynamic traces.
However, the methods above depend heavily on emulation or symbolic execution, which are not suitable for program analysis in large-scale IoT firmware since the emulation requires peripheral devices~\cite{zaddach2014avatar, gustafson2019toward, chen2016towards} and symbolic execution suffers from the problems of path explosion.

\subsection{AST in Source Code Analysis}
{Since the AST can be easily generated from source code, there has been research work proposed to detect source code clone based on AST.
Ira D. Baxter \etal{} proposed to hash ASTs of functions to buckets and compare the ASTs in the same bucket ~\cite{baxter1998clone} to find clones.
Because the method proposed in~\cite{baxter1998clone} is similar to \diaphora{} which hash ASTs, we only perform a comparative evaluation with \diaphora{}.
In addition to the code clone detection, AST is also used in vulnerability extrapolation from source code~\cite{yamaguchi2012generalized,yamaguchi2011vulnerability}. In order to find vulnerable codes that share a similar pattern, Fabian \etal{}~\cite{yamaguchi2012generalized} encoded AST into a vector and utilized the latent semantic analysis~\cite{deerwester1990indexing} to decompose the vector to multiple structural pattern vectors and compute the similarity between these pattern vectors.}
Yusuke Shido~\etal{} proposed an automatic source code summary method with extended Tree-LSTM~\cite{shido2019automatic}.

\section{Conclusion}
In this work, we present \toolname{}, a domain knowledge-enhanced BCSD tool designed to detect homologous vulnerable functions on a broad scale in efficient and accurate manner.
\toolname{} introduces domain knowledge before and after deep learning model-based function encoding to eliminate 
a large proportion of non-homologous functions and score homologous functions higher, separately.
The pre-filtering module makes extensive use of function name information prior to function encoding to accelerate the function encoding.
The function call structure is utilized by the re-ranking module following function encoding to calibrate the encoding similarity scores.
\toolname{} is capable of finding homologous functions rapidly and precisely, according to a comprehensive comparison with existing state-of-the-art research.
Furthermore, \toolname{} manages to find 1,482 vulnerable functions in the real-world firmware bug search experiment with high precision of 91.65\%.
The search results for CVE-2017-13001 demonstrate \toolname{} successfully finds inlined vulnerable code.
\toolname{} can aid in detecting vulnerabilities from large-scale firmware binaries to mitigate the attach damage on IoT devices.

\section{Acknowledgement}
This research was supported in part by National Key R\&D Program of China (Grant No. 2022YFB3103904), the Strategic Priority Research Program of Chinese Academy of Sciences (Grant No. XDC02020100), Joint Fund Cultivation Project of National Natural Science Foundation of China (Grant No. U1636120), the Science and Technology Project of State Grid Corporation of China (Grant No. 5700-202258499A-3-0-ZZ), and National Natural Science Foundation of China (Grant No. U1766215).
Any opinions, findings, and conclusions in this paper are those of the authors and do not necessarily reflect the views of the funding agencies.

\bibliographystyle{ACM-Reference-Format}
\bibliography{ref.bib}
\end{document}